\documentclass[pdflatex,sn-mathphys-num]{sn-jnl}% Math and Physical Sciences Numbered Reference Style
\usepackage{graphicx}%
\usepackage{multirow}%
\usepackage{amsmath,amssymb,amsfonts}%
\usepackage{amsthm}%
\usepackage{mathrsfs}%
\usepackage[title]{appendix}%
\usepackage{xcolor}%
\usepackage{textcomp}%
\usepackage{manyfoot}%
\usepackage{booktabs}%
\usepackage{algorithm}%
\usepackage{algorithmicx}%
\usepackage{algpseudocode}%
\usepackage{listings}%
\usepackage{subcaption}
\usepackage[version=4]{mhchem}
\usepackage{placeins}
\usepackage{makecell}
\usepackage{kpfonts}
\usepackage{lmodern}
\theoremstyle{thmstyleone}%
\theoremstyle{thmstyletwo}%

\theoremstyle{thmstylethree}%

\makeatletter
\newcommand{\beginsupplement}{%
    \setcounter{section}{1}
    \setcounter{subsection}{0}
    \setcounter{table}{0}
    \setcounter{figure}{0}%
    \makeatletter
    \renewcommand{\thesection}{S\arabic{section}}%
    \renewcommand{\thetable}{S\arabic{table}}%
    \renewcommand{\thefigure}{S\arabic{figure}}%
    \renewcommand{\fnum@table}{Supplementary Table~\arabic{table}}%
    \renewcommand{\fnum@figure}{Supplementary Fig.~\arabic{figure}}%
    \makeatother
}

\begin{document}

\title[Article Title]{The Loss Landscape of Powder X-Ray Diffraction-Based Structure Optimization Is Too Rough for Gradient Descent}


% \author*[1,2]{\fnm{First} \sur{Author}}\email{iauthor@gmail.com}



% \affil*[1]{\orgdiv{Department}, \orgname{Organization}, \orgaddress{\street{Street}, \city{City}, \postcode{100190}, \state{State}, \country{Country}}}

% \affil[2]{\orgdiv{Department}, \orgname{Organization}, \orgaddress{\street{Street}, \city{City}, \postcode{10587}, \state{State}, \country{Country}}}

% \affil[3]{\orgdiv{Department}, \orgname{Organization}, \orgaddress{\street{Street}, \city{City}, \postcode{610101}, \state{State}, \country{Country}}}

\author[1]{\fnm{Nofit} \sur{Segal}}\email{nofit@mit.edu}

\author[1]{\fnm{Akshay} \sur{Subramanian}}\email{akshay_s@mit.edu}

\author[2]{\fnm{Mingda} \sur{Li}}\email{mingda@mit.edu}

\author[3]{\fnm{Benjamin Kurt} \sur{Miller}}\email{bkmi@meta.com}

\author*[1]{\fnm{Rafael} \sur{Gómez-Bombarelli}}\email{rafagb@mit.edu}

\affil[1]{\orgdiv{Department of Materials Science and Engineering}, \orgname{Massachusetts Institute of Technology}, \orgaddress{\city{Cambridge}, \state{MA}, \postcode{02139}, \country{USA}}}

\affil[2]{\orgdiv{Department of Nuclear Science and Engineering}, \orgname{Massachusetts Institute of Technology}, \orgaddress{\city{Cambridge}, \state{MA}, \postcode{02139}, \country{USA}}}

\affil[3]{\orgdiv{FAIR}, \orgname{Meta}, \orgaddress{\city{San Francisco}, \state{CA}, \country{USA}}}

\abstract{Solving crystal structures from powder X-ray diffraction (XRD) is a central challenge in materials characterization. In this work, we study the powder XRD-to-structure mapping using gradient descent optimization, with the goal of recovering the correct structure from moderately distorted initial states based solely on XRD similarity. We show that commonly used XRD similarity metrics result in a highly non-convex landscape, complicating direct optimization. Constraining the optimization to the ground-truth crystal family significantly improves recovery, yielding higher match rates and increased mutual information and correlation scores between structural similarity and XRD similarity. Nevertheless, the landscape may remain non-convex along certain symmetry axes. These findings suggest that symmetry-aware inductive biases could play a meaningful role in helping learning models navigate the inverse mapping from diffraction to structure.}


\maketitle

\section{Introduction}\label{intro}
Determining the atomic structure of a crystal from its powder X-ray diffraction (XRD) pattern is a longstanding and central challenge in materials characterization \cite{bragg1914analysis, david2008structure}. The inverse problem of recovering the full three-dimensional crystal structure solely from an XRD pattern is extremely challenging due to the loss of phase information of the scattered waves \textemdash known as the phase problem \cite{hauptman1991phase}\cite[Chapter~9.3, 13.2]{hammond2015basics}. Nevertheless, powder diffraction is widely used for identifying and characterizing crystalline solids. In practice, this is typically achieved by comparing the observed XRD spectrum to a reference database and performing least-squares refinement, known as Rietveld analysis \cite{rietveld1969profile, gates2019powder, allen2006cambridge}. However, this process is highly sensitive to initial parameters \cite{biwer2025spotlight}, and more importantly, it relies on the presence of the correct structure in the database and cannot be used to reconstruct novel or unreported phases.

Experimental phenomena such as preferred orientation, peak overlap, crystal twinning, and instrumental noise further complicate structural determination from powder XRD patterns \cite{holder2019tutorial, chandra1999analysis, szymanski2024integrated}. Moreover, many minerals and metallic alloys exhibit solid solution ranges with very slight lattice shifts, which result in ranges of stoichiometries with nearly the same diffraction pattern \cite[Chapter~10.3.2]{hammond2015basics}\cite{deng2022preparation, wang2024comparison, el2023robust}, making XRD-to-structure mapping a one-to-many problem. Consequently, accurate structure reconstruction typically requires refinement model fitting and domain-specific prior knowledge.

From a computational perspective, structural ambiguity remains even under idealized conditions. Two structures with different compositions can exhibit highly correlated XRD patterns if they share similar symmetry \cite{gu2017sustainable, ayeshamariam2014morphological, ashokkumar2024green}. Moreover, even when stoichiometry is fixed, structures with close though distinct space groups can yield highly similar XRD patterns \cite{david2008structure}. Notably, small distortions in lattice parameters or atomic coordinates can cause discontinuous changes in the diffraction pattern, such as the appearance or disappearance of peaks due to shifting Bragg conditions \cite[Chapter~9]{hammond2015basics}. This introduces a highly non-smooth relationship between structure and XRD signal.

Recently, there has been a surge of interest in crystal structure determination from XRD patterns using generative modeling \cite{guo2025ab, riesel2024crystal, johansen2025decifer, li2025powder, lai2025end, guo2024towards}. A growing body of work applies gradient-based optimization approaches that leverage differentiable physics to refine generated or otherwise-obtained crystal structures by minimizing the difference between simulated and target XRD patterns. For example, \citet{riesel2024crystal} generate crystals conditioned on a given XRD pattern and post-process them using a differentiable XRD simulator to update lattice parameters via gradient descent (GD). \citet{parackal2024identifying} systematically enumerate candidate crystals given composition and space group inputs, and restricts the GD optimization to atomic positions along Wyckoff degrees of freedom. \citet{lee2023creation} create candidate crystals using an evolutionary algorithm, followed by crystals morphing by maximizing the cosine similarity between the XRD patterns.  Outside the powder diffraction setting, GD has further been applied to determine lattice parameters from single-crystal diffraction patterns \cite{gevorkov2019xgandalf}. 

As gradient-based refinement relies on comparing simulated and target diffraction patterns, recent work has also focused on developing more robust XRD-similarity metrics. \citet{otero2024powder} introduced a cross-correlation-based metric that captures equivalence between diffraction patterns while remaining invariant to lattice distortions. Building on this work, \citet{racioppi2025powder} applied the metric to crystal structure prediction from XRD data, jointly optimizing the structure by minimizing both this similarity metric and the structure’s enthalpy. \citet{hernandez2017using} systematically analyzed the sensitivity of different families of similarity metrics under isotropic lattice strain. \citet{li2021spectral} proposed an entropy-based similarity measure for spectra and demonstrated its utility for molecular database retrieval from mass spectrometry data.

% Although these approaches offer automated structure determination pipelines and post-processing tools for generative models of crystalline materials, one must proceed cautiously since the optimization landscape is complicated and non-convex. 
% As another complicating factor, scientists are typically interested in the subset of structures that are local minima of the quantum-mechanical energy function. Mapping powder XRD diffraction signals to crystal structures is therefore a multi-objective optimization problem.

In this work, we explore the powder XRD-to-structure mapping through the lens of GD optimization. The goal is to recover correct structures based solely on XRD similarity from moderately deviated states. We ask whether the XRD  landscape is locally smooth enough for GD to guide us back to the correct configuration. Inspired by experimentally observed symmetry-breaking effects such as thermal expansion from lattice vibrations and thermal fluctuations, we introduce two types of distortions: random lattice distortions and uncorrelated atomic displacements \cite{biwer2025spotlight, cannelli2022atomic, brivio2015lattice, delgado2013effects}. These distortions resemble crystal structures predicted by generative models, which often produce nearly correct geometries but with imperfect symmetry \cite{jiao2024space, levy2025symmcd, kazeev2025wyckoff}. Through this study, we examine the challenge of "the last mile" in structure elucidation from XRD. 

We find that mapping XRD patterns to crystal structures is challenging because high diffraction agreement, as currently measured in literature, does not ensure structural accuracy. We show that commonly used XRD similarity measures, such as cosine similarity, mean squared error (MSE), and entropy similarity, are sensitive to both lattice and coordinate noise distortions, and optimization between distorted structures and ground-truth XRD diffraction can become trapped in local minima. We explore an alternative strategy that enforces lattice constraints, highlighting the role of symmetry in connecting XRD to the crystal structure. 

%%%%%%%%%%%%%%%%%%%%%%%%%%%%%%%%%%%%%%%%%%%%%%%%%%%%%%%%%%%%
\section{Method}
\label{sec:methods}
\begin{figure}[ht]
\centering
\begin{tabular}{ccc}
 \includegraphics[width=0.20\textwidth]{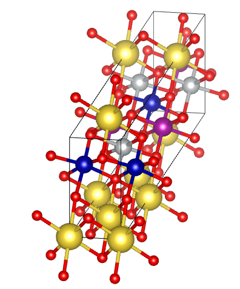} &
 \includegraphics[width=0.20\textwidth]{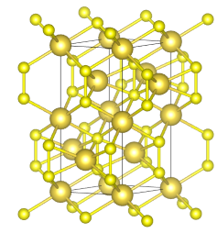} &
\includegraphics[width=0.20\textwidth]{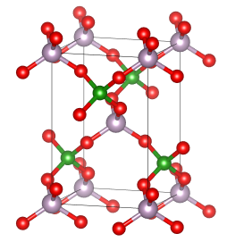} \\
\textit{\ce{Na3MnCoNiO6};  Cm (8)} &
\textit{\ce{NaS}; P6$_3$/mmc (194)} &
\textit{\ce{BPO4}; I$\bar{4}$ (82)} \\[0.8em]

\includegraphics[width=0.20\textwidth]{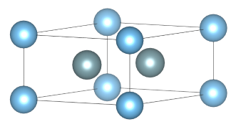} &
\includegraphics[width=0.20\textwidth]{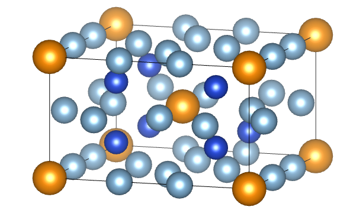} &
\includegraphics[width=0.20\textwidth]{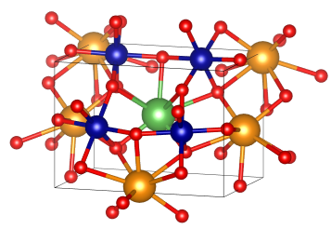} \\
\textit{\ce{U2Ti}; P6/mmm (191)} &
\textit{\ce{Nd(Al2Cu)4}; I4/mmm (139)} &
\textit{\ce{LaNd3Cr4O12}; Pm (6)} \\[0.8em]

\includegraphics[width=0.20\textwidth]{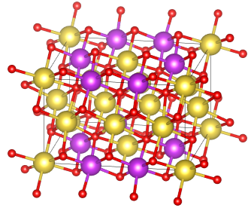} &
\includegraphics[width=0.20\textwidth]{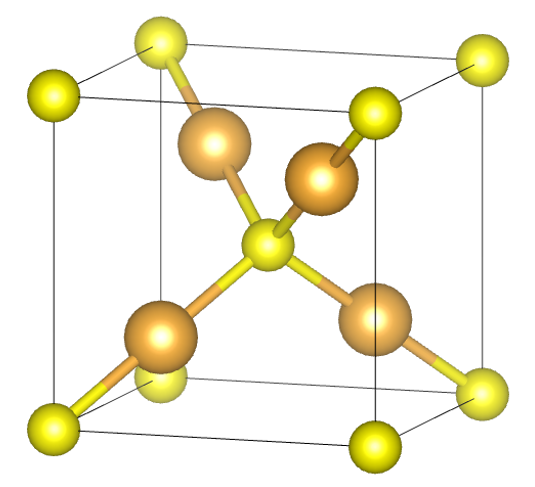} & 
\includegraphics[width=0.20\textwidth]{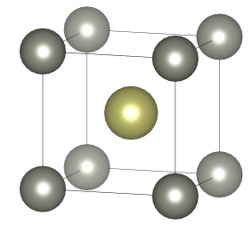} \\
\textit{\ce{Na2BiO3}; C2/m (12)} &
\textit{\ce{Au2S};  Pn$\bar{3}$m (224)} &
\textit{\ce{HfZn}; Pm$\bar{3}$m (221)} \\[0.8em]

\includegraphics[width=0.20\textwidth]{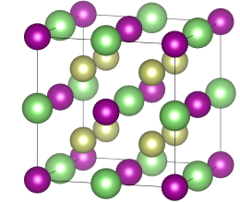} & & \\
\textit{\ce{LiMnIr2}; Fm$\bar{3}$m (225)} & & \\
\end{tabular}
\caption{\textbf{Crystal structures used in this study.} Each structure is labeled with its chemical formula, space group symbol, and space group number.}
\label{apndx:structs}
\end{figure}

We selected 10 structures from the MP20 dataset \cite{xie2021crystal}, a collection of small, inorganic, thermodynamically (meta)stable structures from the Materials Project \cite{jain2013commentary}. We selected the structures according to the most common space groups (see Figure~\ref{apndx:structs}), which span a range of crystal symmetries:

\noindent
$P6/mmm,\ Pn\bar{3}m,\ I\bar{4},\ Cm,\ I_4/mmm,\ Fm\bar{3}m,\ C2/m,\ P6_3/mmc,\ Pm\bar{3}m,\ Pm$.

For each structure, we generated 50 distorted versions using two noise models:

\subsection{Lattice Noise}
Random lattice distortions were applied via strain tensors \cite{szymanski2024integrated}. These modifications alter the cell while keeping fractional atomic coordinates fixed. Each distorted structure was generated by randomly sampling the entries of a strain tensor and applying it to the ground-truth lattice matrix.  Let $\mathbf{L}\in\mathbb{R}^{3\times 3}$ be the ground-truth lattice matrix (columns are the lattice vectors).
We noise the lattice by a random deformation matrix $\mathbf{S}\in\mathbb{R}^{3\times 3}$,
\[
\tilde{\mathbf{L}}=\mathbf{S}\,\mathbf{L},\qquad
\mathbf{S}=\begin{pmatrix}
s_{11} & s_{12} & s_{13}\\
s_{21} & s_{22} & s_{23}\\
s_{31} & s_{32} & s_{33}
\end{pmatrix},
\]
with no change to the atomic fractional coordinates. For a noise level $\sigma_l>0$, the entries of $\mathbf{S}$ are sampled as
\begin{align*}
s_{ii} &\sim \mathrm{Unif}\ \big(1-\sigma_l,\;1+\sigma_l\big), && i\in\{1,2,3\},\\
s_{ij} &\sim \mathrm{Unif}\ \big(-\sigma_l,\;\sigma_l\big), && i\neq j.
\end{align*}
Thus, diagonal entries produce uniaxial expansion/compression, whereas off-diagonal entries induce shear.
By construction, these distortions do not preserve crystal symmetry and can introduce diverse deformation modes. If $\boldsymbol{f}$ denotes a fractional coordinate, then the
Cartesian position changes from $\boldsymbol{x}=\mathbf{L}\boldsymbol{f}$ to $\tilde{\boldsymbol{x}}=\tilde{\mathbf{L}}\boldsymbol{f}$ with $\boldsymbol{f}$ unchanged.

\subsection{Coordinate Noise}
Independent, uncorrelated positional perturbations were applied to each atom by adding Gaussian-distributed noise to its fractional coordinates. Let $x_i^{(n)} \in [0,1)^3$ be the fractional coordinates of atom $n$.
For a noise scale $\sigma_c>0$, we draw i.i.d.\ perturbations
\[
\varepsilon_n \sim \mathcal{N}\!\big(0,\;\sigma_c^2 I_3\big), \qquad n=1,\dots,N,
\]
and set the noisy fractional coordinates to
\[
x_f^{(n)} \;=\; w\!\big(x_i^{(n)}+\varepsilon_n\big), \qquad
w(u)\;=\;u-\lfloor u\rfloor \in [0,1)^3,
\]
where $\lfloor u\rfloor$ applies the floor function componentwise.
Equivalently, $x_f^{(n)} \equiv x_i^{(n)}+\varepsilon_n \pmod{1}$ (elementwise), i.e., on the 3-torus $\mathbb{T}^3=\mathbb{R}^3/\mathbb{Z}^3$.

After applying noise, we attempted to recover the ground-truth structure via gradient-based optimization using StructSnap, a differentiable XRD simulator \cite{riesel2024crystal}.
We compute diffraction patterns from the structure factor contributions of each atomic site, following Bragg’s law and the kinematic scattering model. The resulting pattern is a 2D tensor of $2\theta$ angles and intensities.

\subsection{Optimization} We refine either the lattice parameters or atomic coordinates in accordance with the noise applied. The structure is passed through a differentiable diffraction pipeline to produce a simulated pattern, which is then compared to the ground-truth pattern using a chosen loss function. Given distorted-state XRD $\hat{\mathbf{x}}$ and ground-truth XRD $\mathbf{x}$ (see~\ref{appndx:xrd_details}), we minimize the negative cosine similarity, MSE loss, or negative entropy similarity \cite{li2021spectral},  which are defined respectively:
\begin{align*}
\mathcal{L}_{\mathrm{cos}}
&= \frac{1}{N}\sum_{i=1}^N
\left(
-\,\frac{\mathbf{x}^{(i)} \cdot \hat{\mathbf{x}}^{(i)}}{\lVert \mathbf{x}^{(i)} \rVert_2 \, \lVert \hat{\mathbf{x}}^{(i)} \rVert_2}
\right),
&\quad&
\mathcal{L}_{\mathrm{MSE}}
= \frac{1}{N}\sum_{i=1}^N \left\lVert \mathbf{x}^{(i)} - \hat{\mathbf{x}}^{(i)} \right\rVert_2^{2},
\\[6pt]
\mathcal{L}_{\mathrm{Entropy}}
&= -\frac{1}{N}\sum_{i=1}^N \bigg( 1-\frac{2S_{\hat{x_i}x_i}-S_{x_i}-S_{\hat{x_i}}}{\text{log} 4}\bigg),
&\quad&
\end{align*}

where \( S \) denotes the Shannon entropy.
Gradients of the loss are backpropagated to update structural parameters using PyTorch’s autograd. 

\subsection{Symmetry Constraints}
\label{subsec:method_sym_const} We examine the effect of symmetry-based constraints in the lattice-noise case by enforcing the ground-truth crystal family during optimization. At inference time, the ground truth information will not be available. However, previous works have shown strong performance in predicting the crystal family \cite{suzuki2020symmetry, cao2024simxrd, lee2022powder, zhang2024crystallographic, bin2025simxrd}, as well as lattice parameters \cite{habershon2004powder, chitturi2021automated, dong2021deep} and space groups \cite{schopmans2023neural, bin2025simxrd, oviedo2019fast} from XRD patterns. 

Using projected optimization, each gradient step is followed by projection onto the constrained values. 

Let $\theta = (a,b,c,\alpha,\beta,\gamma)$ be the lattice parameters. Simple constrained gradient descent on a proposed crystal $\hat{\mathcal{X}}$ using user-selected loss $\mathcal{L} \in \{\mathcal{L}_{\cos}, \mathcal{L}_{\textrm{MSE}} , \mathcal{L}_{\textrm{entropy}}\}$ and symmetry projection operator $\mathcal{P}$ would be:
\begin{align*}
    \theta^{0} = \mathcal{P}(\theta^{\textrm{init}}) , & &
    \theta^{k+1} &= \mathcal{P} \left( \theta^{k} - \nabla_{\theta^{k}} \mathcal{L}\left( \mathrm{xrd}\left(\hat{\mathcal{X}}(\theta^{k})\right), \mathrm{xrd}_0 \right) \right),
\end{align*}
with the iterations repeating until some convergence criterion is satisfied. Note that $\theta^{\textrm{init}}$ are initial lattice parameters from our prediction, model, or, in this case, distorted ground-truth; $\hat{\mathcal{X}}(\theta^{k})$ is the representation of our proposed crystal structure, which depends on current lattice parameters $\theta^{k}$; $\textrm{xrd}$ is a map from crystal to computed powder x-ray diffraction pattern; $\textrm{xrd}_0$ is the xrd pattern of the reference we aim to recover. The projection operator $\mathcal{P}$ is defined by relevant crystal family, e.g., 
\begin{equation*}
    \mathcal{P}_{\mathrm{cubic}}(a,b,c,\alpha,\beta,\gamma)
    = \big(\bar{a},\bar{a},\bar{a},90^\circ,90^\circ,90^\circ\big),
    \quad
    \bar{a} = \tfrac{a+b+c}{3}.
\end{equation*}
Thus, $a$, $b$, and $c$ are first updated independently according to their gradients, then set to the mean value $\bar{a}$, while angles are fixed to $90^\circ$. Similar projectors are defined for the remaining crystal families (see~\ref{app:sym_op}).

Recovery performance is assessed primarily using Match Rate, the fraction of optimized structures that are identified as structurally equivalent to the ground-truth by \texttt{StructureMatcher}~\cite{ong2013python}, considering lattice, atomic positions, and symmetry.  The tolerances used are 0.1 for lattice, 0.2 for atomic site positions, and 5 degrees for angles. Additionally, we assess recovery using the Average Minimum Distance (AMD), which serves as a complementary metric for quantifying structural similarity between periodic crystals \cite{widdowson2022resolving}. While match rate provides a binary classification of whether two structures are equivalent within a given tolerance, AMD offers a continuous measure that compares the distributions of interatomic distances.

%%%%%%%%%%%%%%%%%%%%%%%%%%%%%%%%%%%%%%%%%%%%%%%%%%%%%%%%%%%%

\section{Results}
\begin{figure}
    \centering
    \begin{subfigure}[t]{0.49\textwidth}
        \centering
        \includegraphics[width=\textwidth]{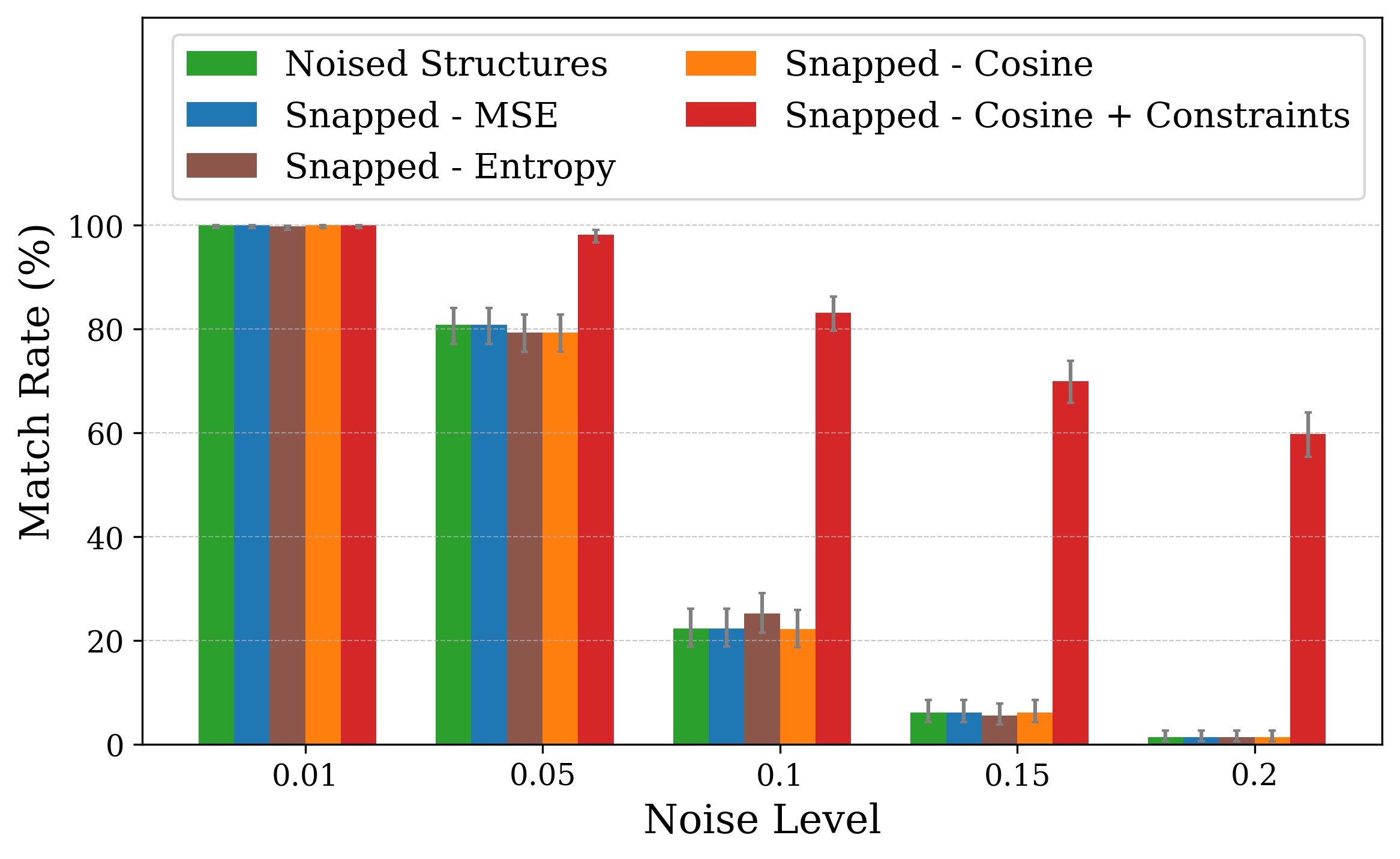}
        \caption{Lattice noise}
        \label{fig:match_lattice}
    \end{subfigure}
    \hfill
    \begin{subfigure}[t]{0.49\textwidth}
        \centering
        \includegraphics[width=\textwidth]{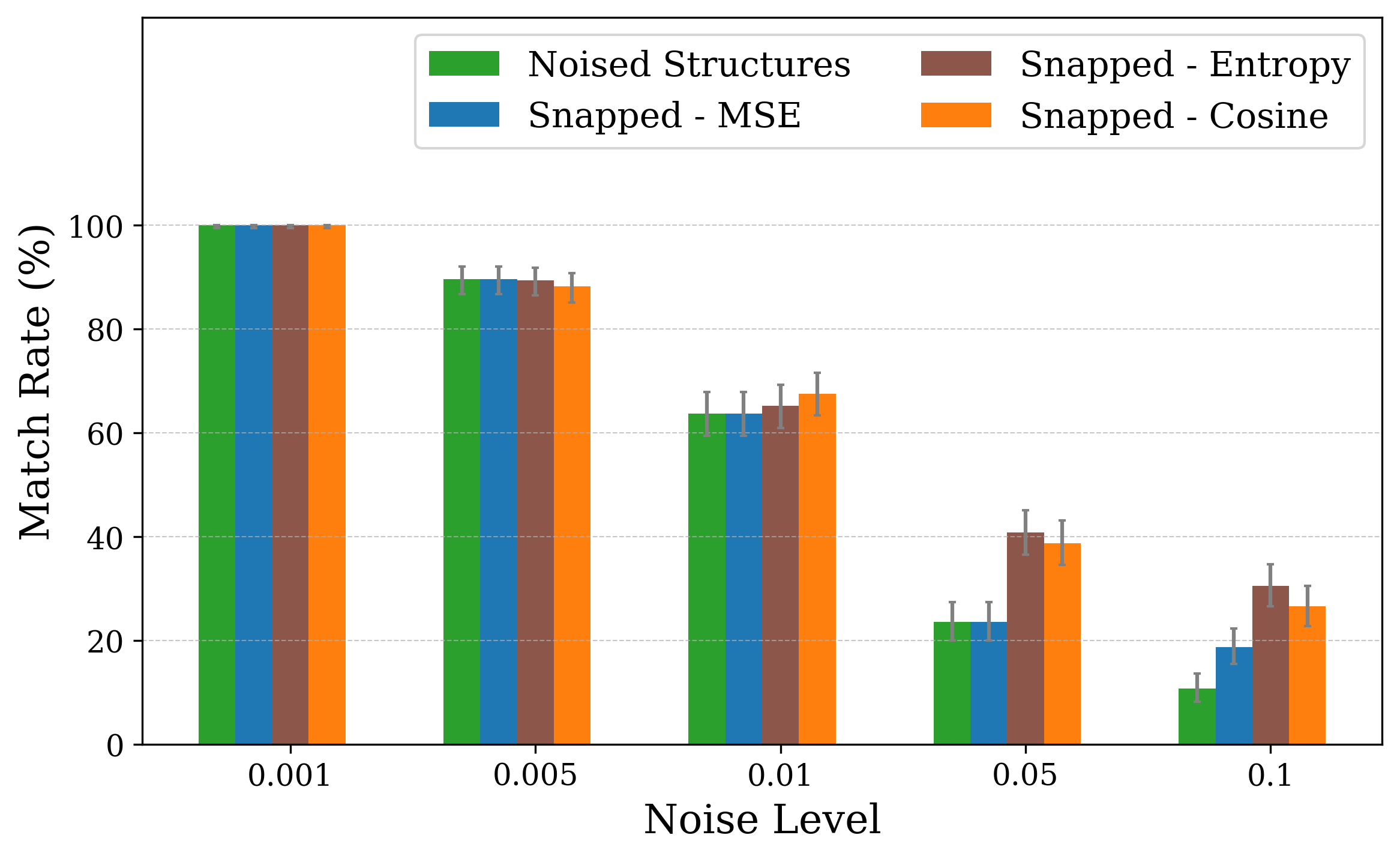}
        \caption{Coordinate noise}
        \label{fig:match_coords}
    \end{subfigure}
    \caption{\textbf{Results of XRD-based optimization under two types of structural noise.} Crystal structures were optimized with respect to XRD similarity metrics using the \texttt{snap} method \cite{riesel2024crystal}, which struggles to recover the correct structure under both lattice and coordinate perturbations. The plots show match rates computed with \texttt{StructureMatcher} ($\text{ltol}=0.1, \ \text{stol}=0.2, \ \text{angle\_{tol}}= 5^\circ$) under random lattice (a) and coordinate (b) perturbations. Error bars represent 95\% Jeffreys binomial credible intervals \cite{brown2001interval}. For lattice distortions, incorporating symmetry constraints significantly improves robustness, even at high noise levels.}
    \label{fig:match_comparison}
\end{figure}

We performed optimization on distorted crystal structures across a range of noise types and levels. For each condition, 50 distorted versions were generated for each of 10 ground truth structures, yielding 500 distorted inputs per noise setting. The 50 variations per structure enable a statistical view. We report the match rate of the optimized structures to the ground truth in Figure~\ref{fig:match_comparison}.

Figure~\ref{fig:match_comparison} illustrates match rates obtained through crystal structure optimization w.r.t XRD similarity metrics. In particular, for lattice distortions, the largest drop occurs between noise levels of 0.05 and 0.1. While 0.1 is the lattice tolerance threshold for matching (see Section~\ref{sec:methods}), the noise level defines the maximum possible strain sampled, so lattice lengths remain within the tolerable range. Using either cosine similarity or MSE as the similarity objective makes little to no difference in performance.

\subsection{Symmetry Constraints: Strengths and Limitations}\label{subsec:res_sym_const}
Incorporating symmetry-based constraints during XRD-based optimization notably improves robustness to lattice noise for many structures in this study, as shown in Figure~\ref{fig:match_lattice} by the higher match rates achieved when constraints are applied. Our constraints (see subsection~\ref{subsec:method_sym_const}) project updates back into the correct crystal family at each optimization step, thereby guiding the search along a reduced-dimensionality symmetry-consistent path and helping the optimizer avoid local minima unrelated to the desired symmetry. 

\begin{figure*}[ht]
    \centering
    \begin{subfigure}[t]{0.58\textwidth}
        \centering
        \includegraphics[width=\textwidth]{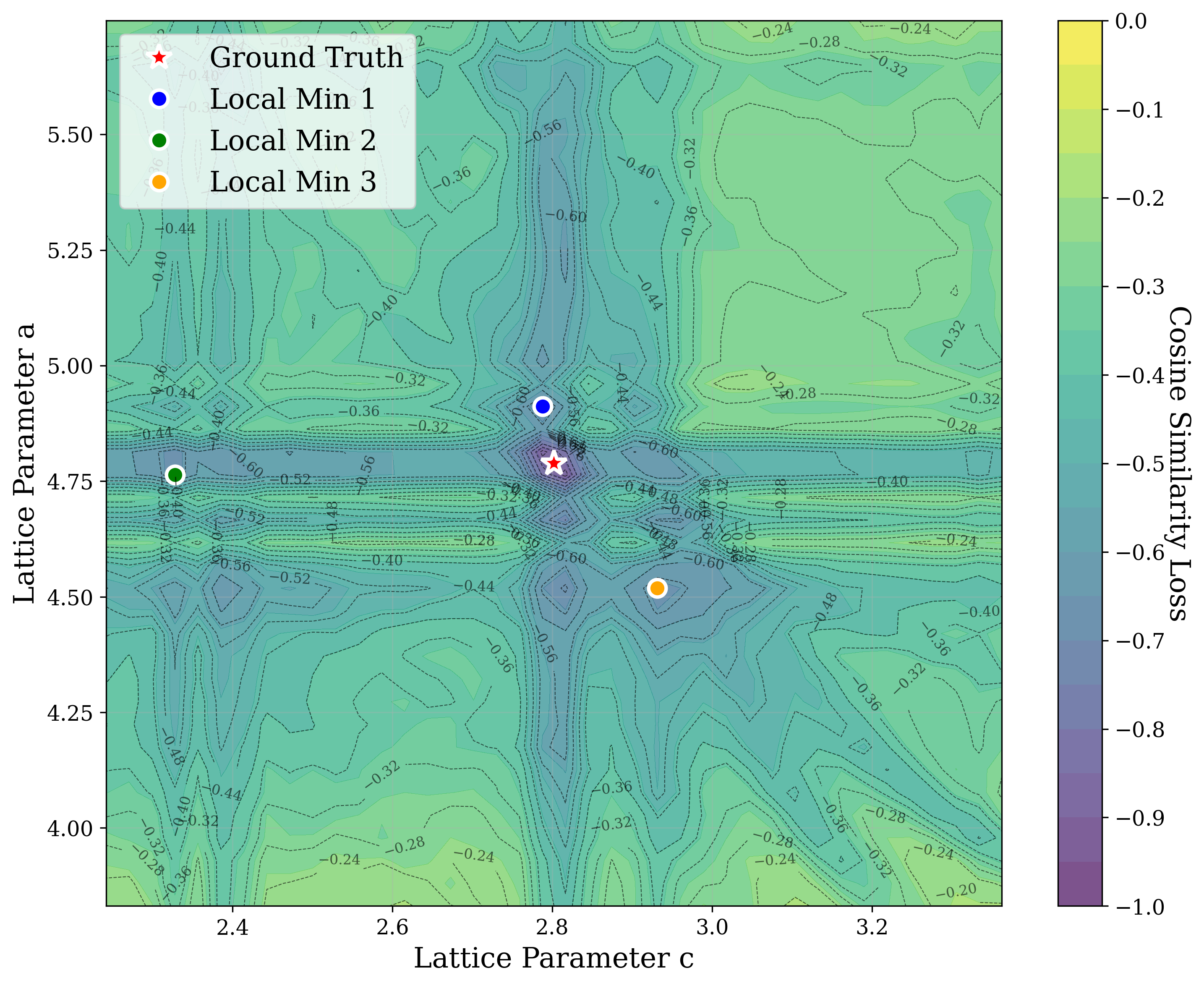}
        \caption{}
    \end{subfigure}
    \hfill
    \begin{subfigure}[t]{0.412\textwidth}
        \centering
        \includegraphics[width=\textwidth]{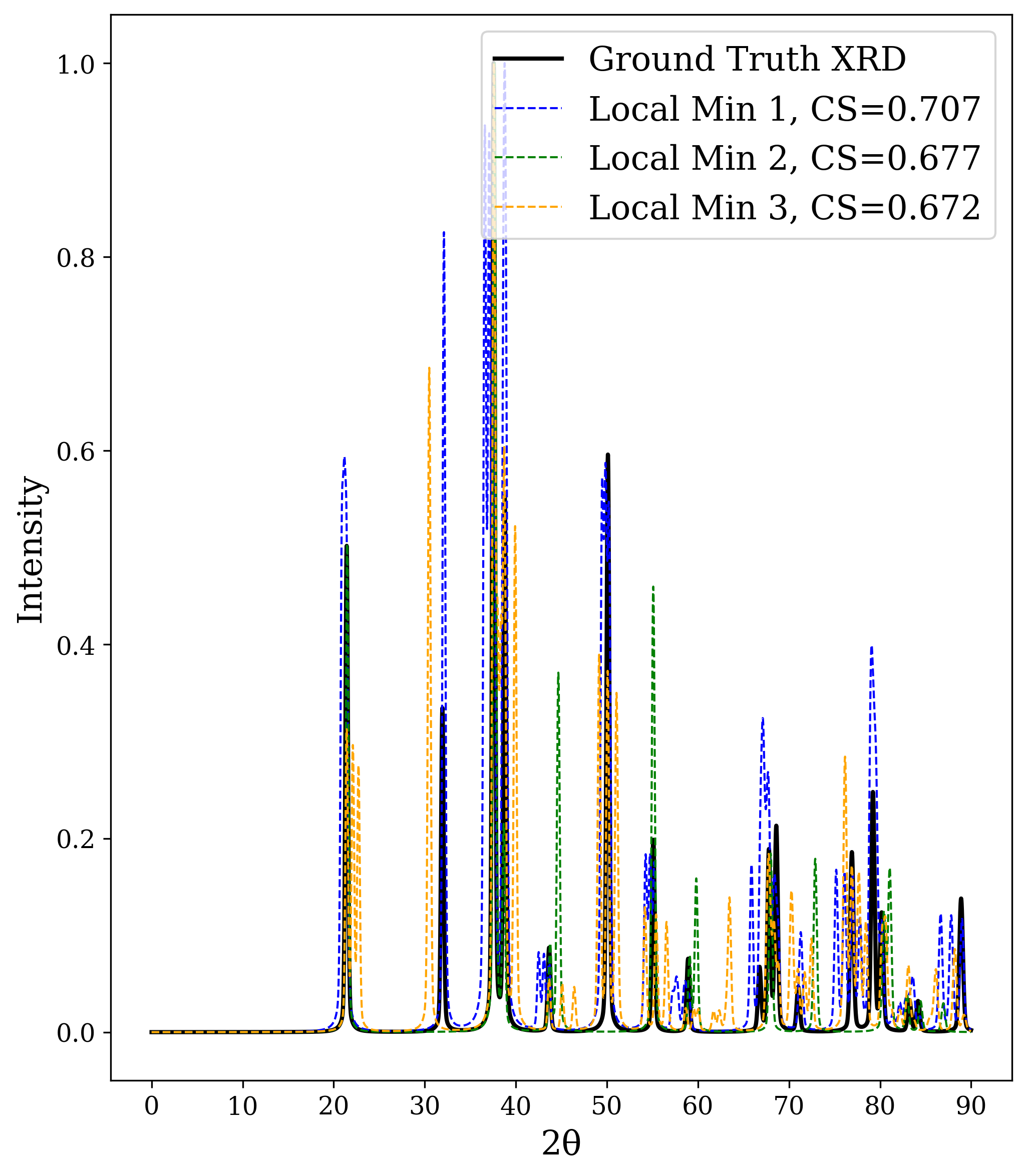}
        \caption{}
    \end{subfigure}
    \caption{\textbf{2D landscape of XRD cosine similarity (CS) loss as a function of lattice parameters $a$ and $c$ of \ce{U2Ti} structure, illustrating the presence of multiple local minima.} (a) Cosine similarity loss topographic map showing non-convex behavior with several local minima. (b) XRD patterns for the structures corresponding to the marked local minima: all exhibit reasonably high cosine similarity to the ground truth pattern despite having different lattice parameters.}
    \label{fig:convex}
\end{figure*}

\subsubsection{Roughness of Simplified Loss Landscape Cross Sections}
\begin{figure*}[ht]
    \centering
    \begin{subfigure}[t]{0.48\textwidth}
        \centering
        \includegraphics[width=\textwidth]{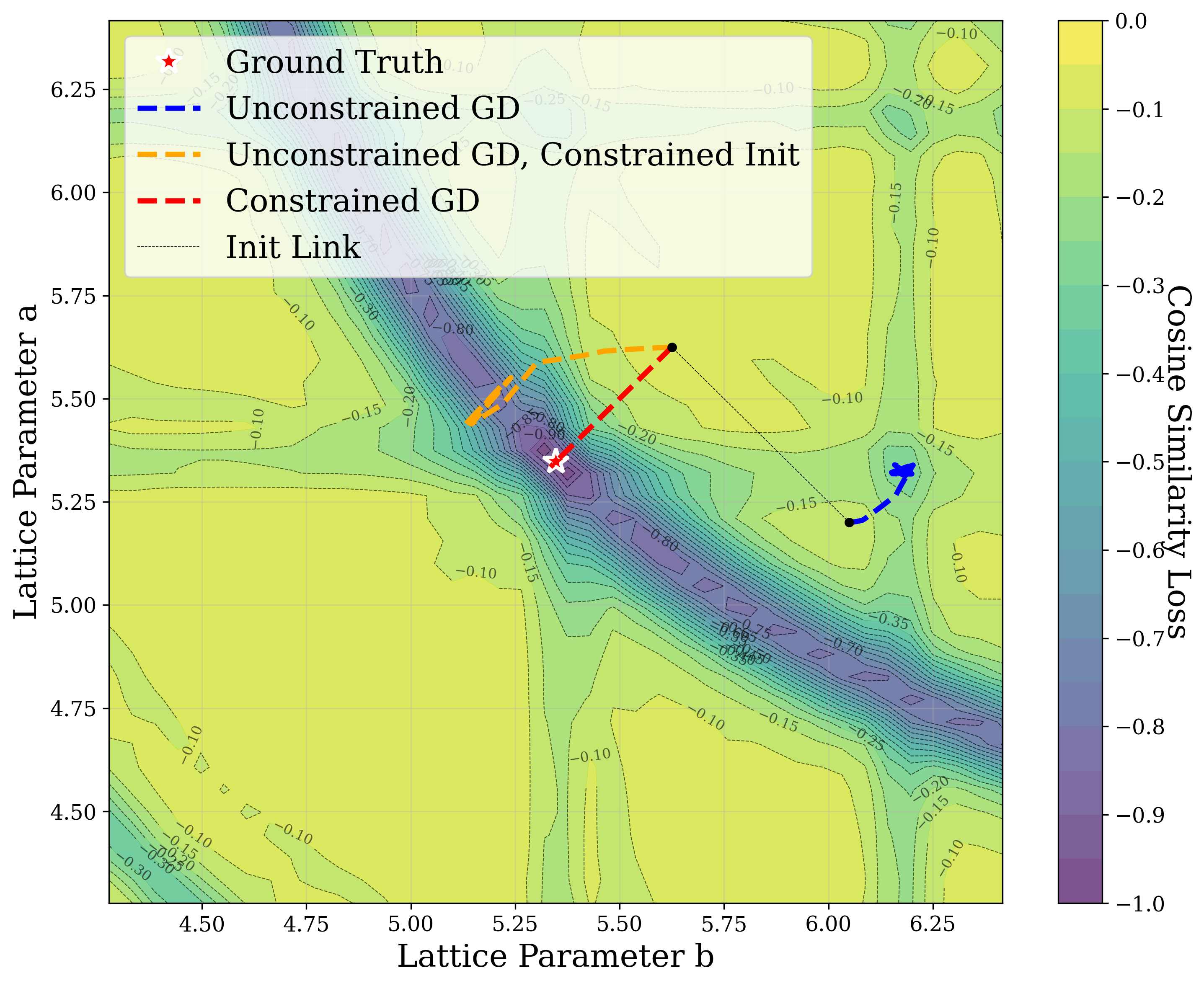}
        \caption{}
        \label{subfig:path_a}
    \end{subfigure}
    \hfill
    \begin{subfigure}[t]{0.48\textwidth}
        \centering
        \includegraphics[width=\textwidth]{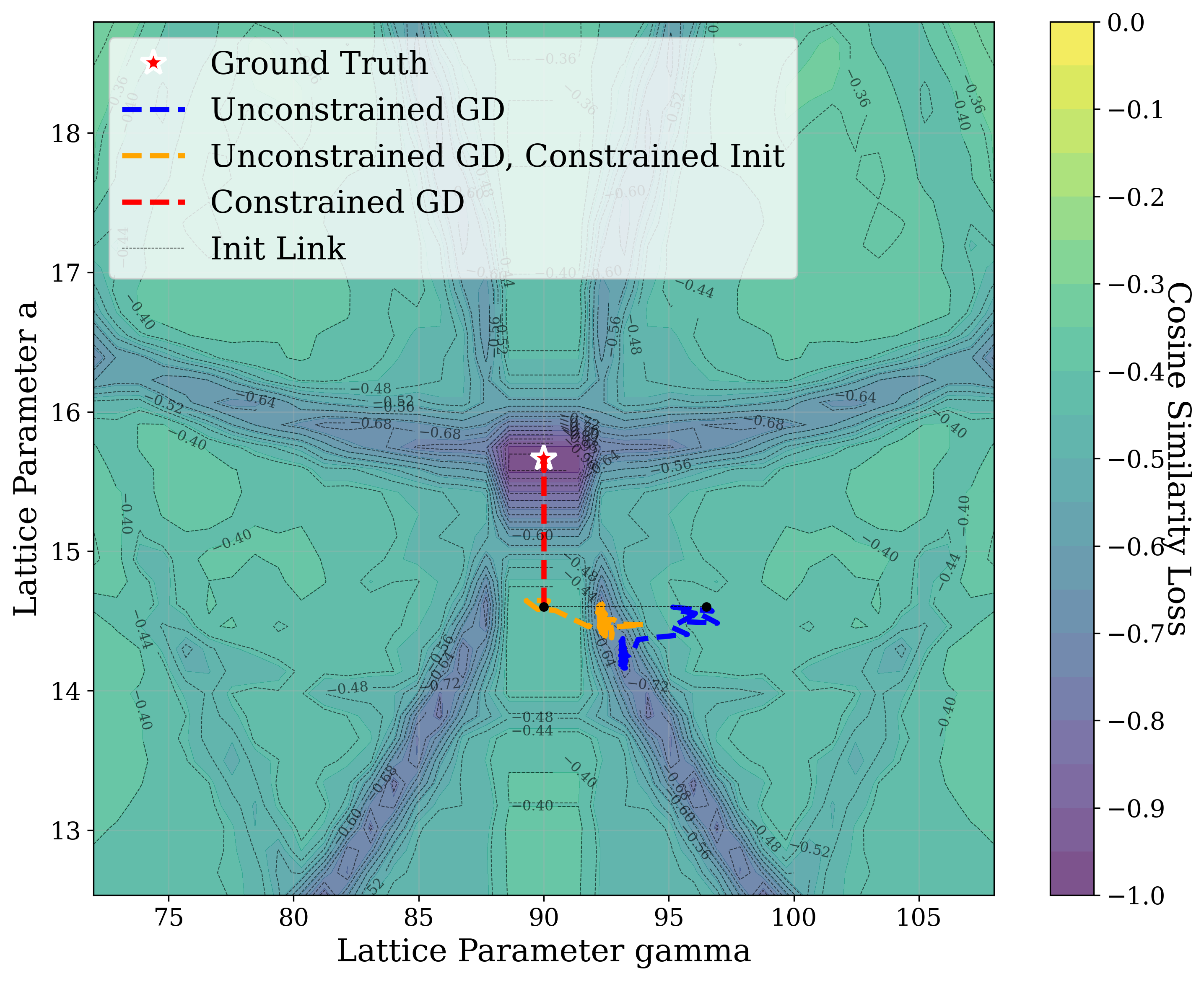}
        \caption{}
        \label{subfig:path_b}
    \end{subfigure}
    \caption{\textbf{2D landscape of XRD cosine similarity (CS) loss as a function of lattice parameters, with simulated optimization paths for XRD-based gradient descent (GD)}. Unconstrained GD, unconstrained GD with a constrained initialization, and fully constrained GD. Unconstrained GD converges to some local minima, even with constrained initialization, whereas constrained GD reaches the ground truth. (a) Lattice parameters $a$ and $b$ of cubic \ce{Au2S} are perturbed. (b) Lattice parameters $a$ and $\gamma$ of monoclinic \ce{Na3MnCoNiO6} are perturbed.}
    \label{fig:convex_paths}
\end{figure*}

\begin{figure}[ht]
\centering
\begin{tabular}{cc}
\includegraphics[width=0.44\textwidth]{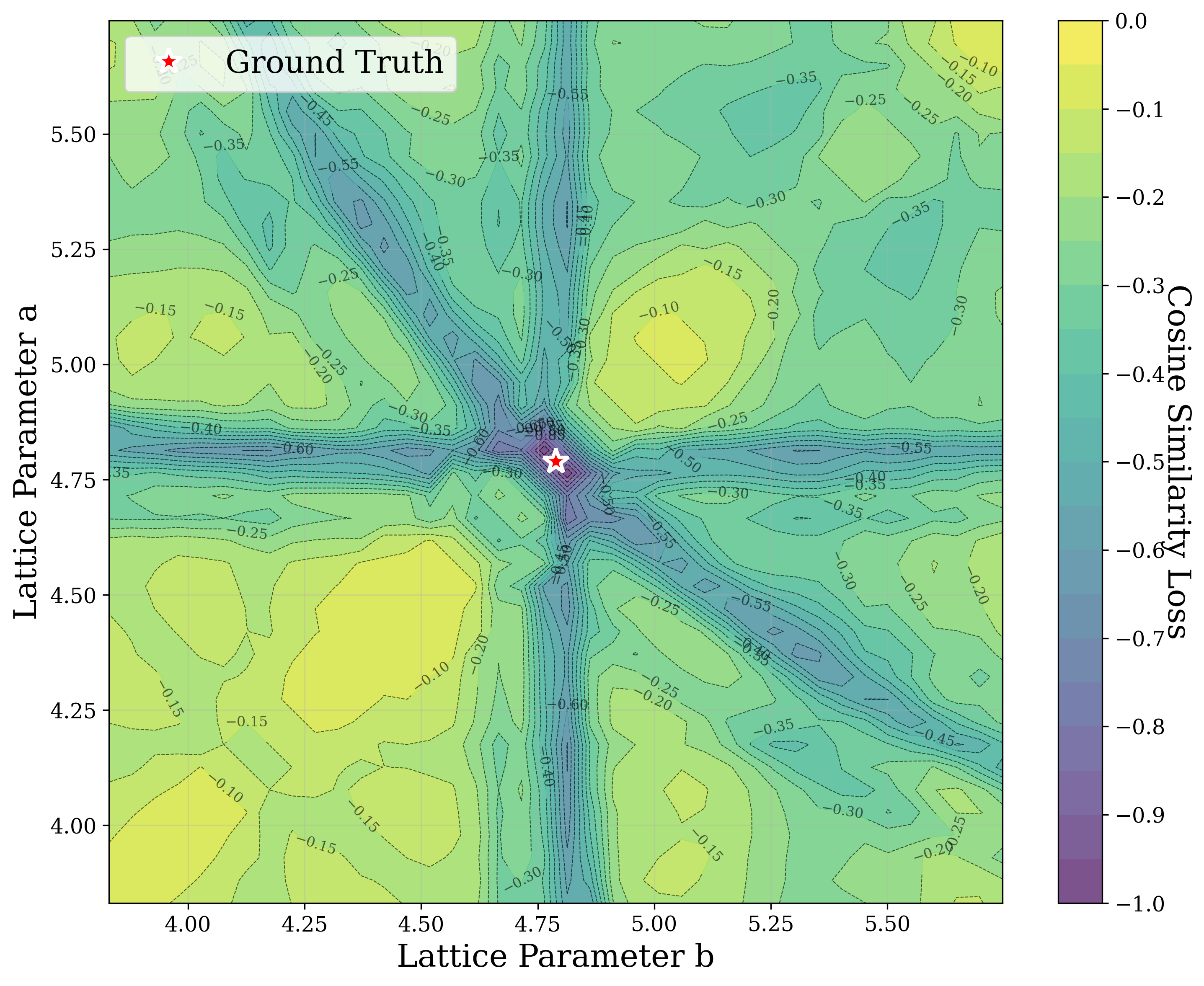} &
\includegraphics[width=0.44\textwidth]{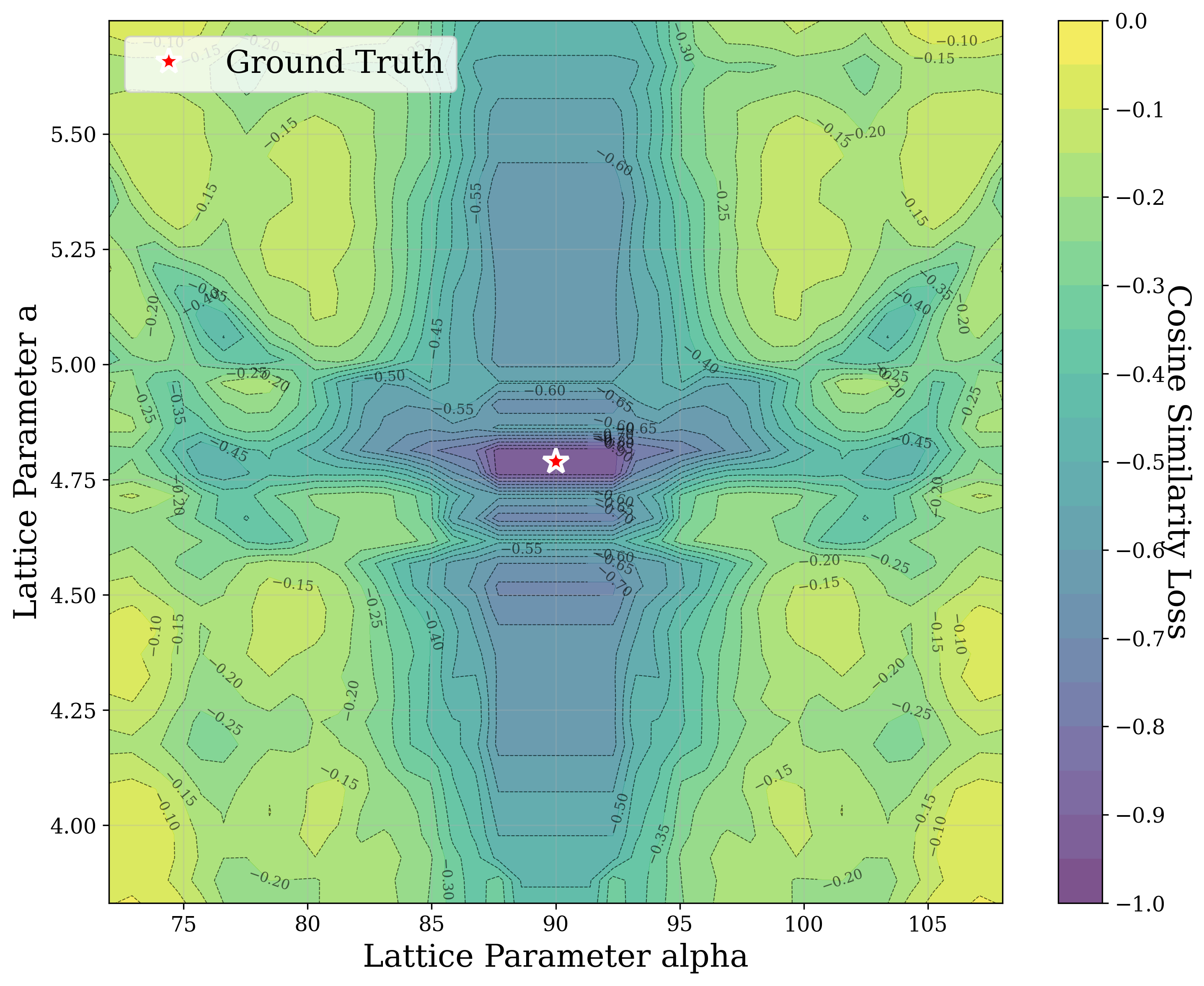} \\
\textit{$a, b$} & \textit{$a, \alpha$} \\[0.8em]

\includegraphics[width=0.44\textwidth]{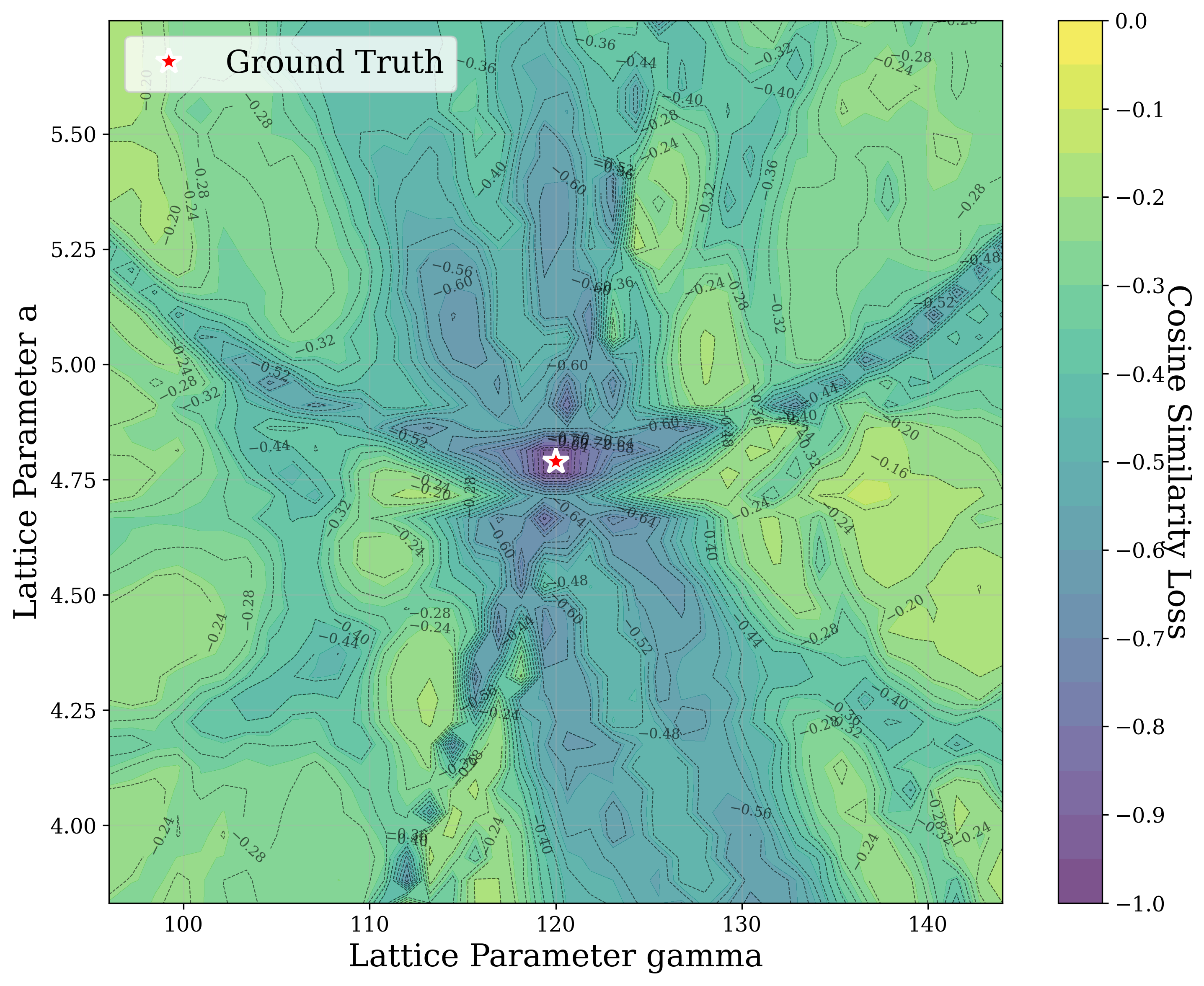} &
\includegraphics[width=0.44\textwidth]{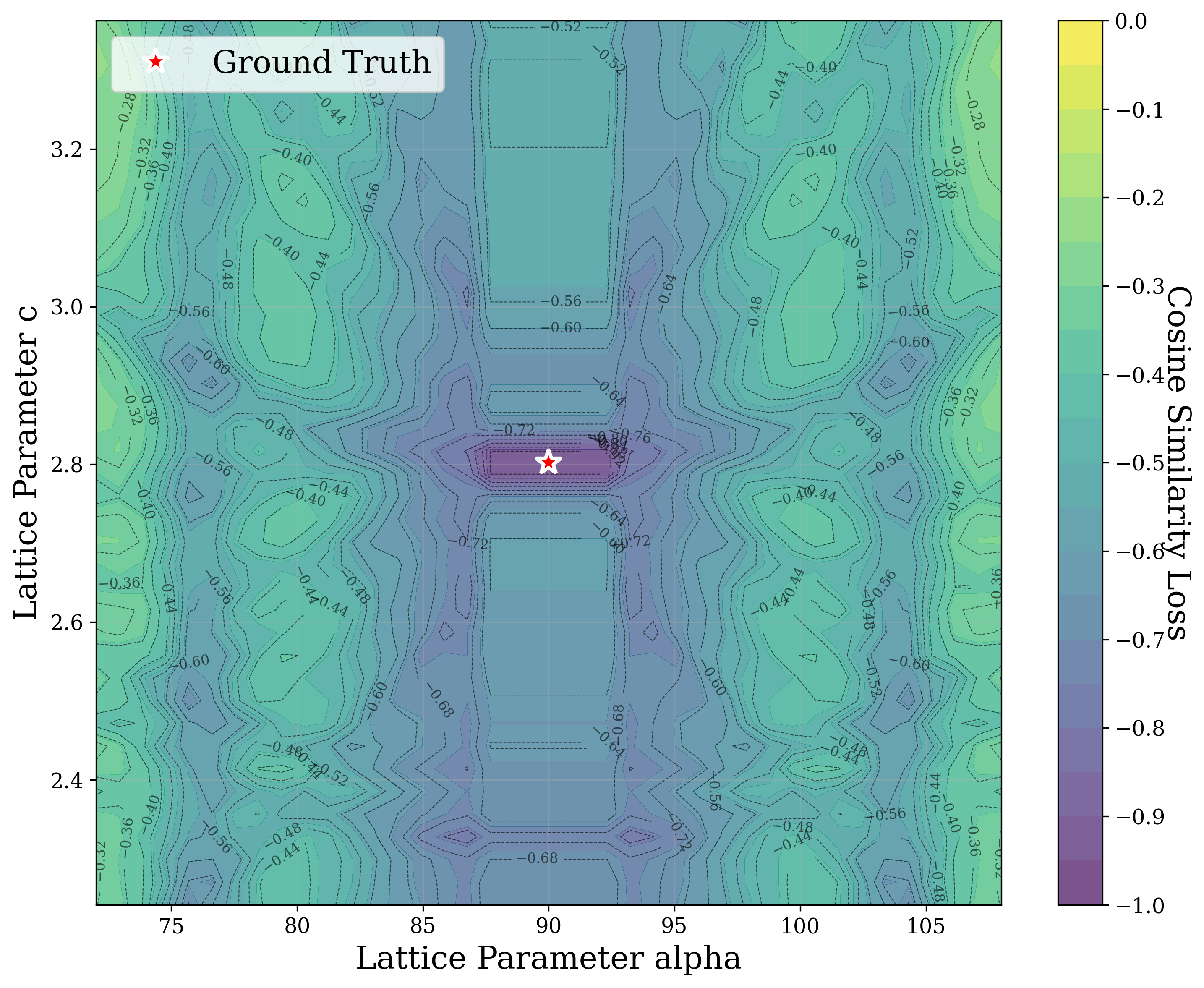} \\
\textit{$a, \gamma$} & \textit{$c, \alpha$} \\[0.8em]

\includegraphics[width=0.44\textwidth]{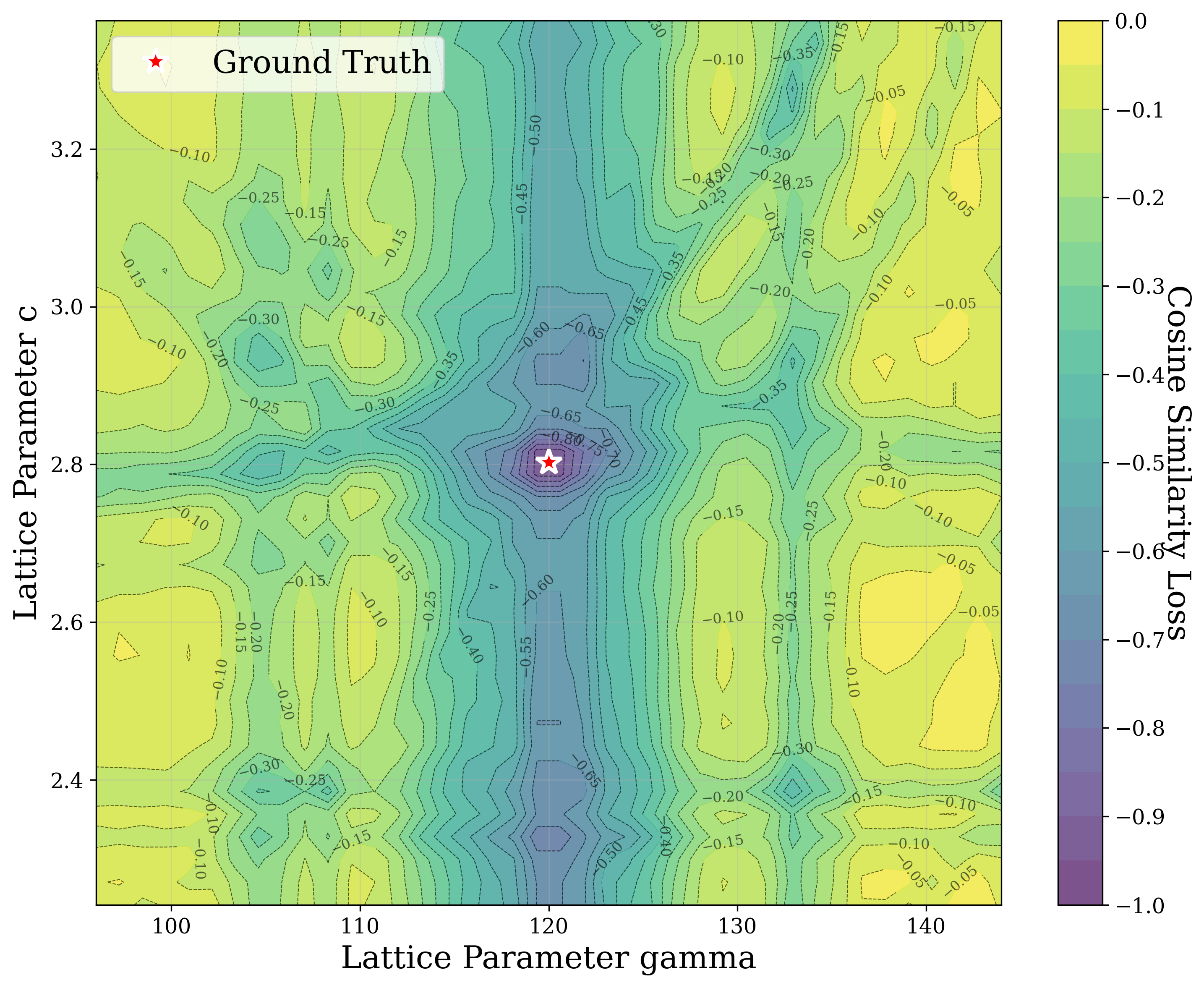} & \\
\textit{$c, \gamma$} & \\
\end{tabular}

\caption{\textbf{All 2D landscape slices of XRD cosine similarity of \ce{U2Ti}.}
Each slice is labeled with the distorted parameters. Due to the hexagonal symmetry of the structure, some slices are redundant and thus obscured.}
\label{apndx:convex_all_slices}
\end{figure}

Figures~\ref{fig:convex}, \ref{fig:convex_paths}, and \ref{apndx:convex_all_slices} illustrate the non-convex nature of the XRD-based loss landscape with respect to the lattice parameters, shown through 2D cross-sections of the optimization surface. These plots represent simplified views of the underlying optimization process, which, in the case of a distorted lattice, occurs in a six-dimensional space corresponding to the six lattice parameters.

In Figure~\ref{fig:convex}, we distort the lattice parameters $a$ and $c$ of \ce{U2Ti}, a hexagonal structure of space group \textit{P6/mmm} (No. 191), and compute the cosine similarity loss between distorted structures' spectra and the ground truth. The resulting contour map reveals multiple deep local minima, indicating the optimizer's potential to get trapped in suboptimal solutions. The three most prominent local minima are highlighted, and their corresponding XRD patterns are shown in the right panel. Despite their structural deviation from the true lattice parameters, the patterns show high cosine similarity to the ground truth due to subtle shifts and peak splittings that preserve the overall spectral profile. 

Figure~\ref{fig:convex_paths} provides two illustrative examples demonstrating how symmetry constraints can facilitate correct structure reconstruction. In these cases, only a two-dimensional slice of the optimization landscape is visualized for clarity. This is a simplification of the full optimization process, which, for the distorted lattice case, occurs in a six-dimensional space. We show simulated GD trajectories for two representative structures under three settings: (i) unconstrained GD, (ii) unconstrained GD initialized at a constrained point, and (iii) fully constrained GD. 

In Figure~\ref{subfig:path_a}, we perturb the lattice parameters $a$ and $\alpha$ of \ce{Au2S}, a cubic structure with space group \textit{Pn$\bar{3}$m} (No.~224). The unconstrained GD trajectory converges to a distant local minimum, whereas unconstrained GD with constrained initialization at $a=b$ reaches a nearby local minimum. By contrast, the fully constrained GD trajectory successfully recovers the ground truth.

In Figure~\ref{subfig:path_b}, we distort the lattice parameters $a$ and $\gamma$ of \ce{Na3MnCoNiO6}, a monoclinic structure with space group \textit{Cm} (No.~8). Similarly, the unconstrained GD trajectory, even when initialized at a constrained point with $\gamma=90^\circ$, converges to a local minimum, while the constrained GD trajectory, which enforces $\gamma=90^\circ$ throughout optimization, reaches the ground truth. 

The initialization points for both cases were chosen for illustration, though points for multiple regions would yield similar behavior. These visualizations demonstrate how symmetry-constrained XRD-based optimization can better reach the correct phase, yielding higher match rates than its unconstrained counterpart (Figure~\ref{fig:match_comparison}). This highlights the value of incorporating symmetry constraints into learning schemes that navigate the structure-to-XRD mapping.

Notably, in Figure~\ref{apndx:convex_all_slices}, we observe fluctuations that pose challenges for symmetry-constrained GD along symmetry axes such as $a=b$ and $\alpha=90^\circ$. While fluctuations along $a=b$ are pronounced, those along $\alpha=90^\circ$ are comparatively shallow and may be mitigated through techniques such as momentum \cite{qian1999momentum} or regularization, which were not studied in this work. Although symmetry constraints generally improve refinement performance, the landscape visualized here highlights that XRD-based GD may remain sensitive to initialization and prone to local minima in some cases.

\subsubsection{Structure Average Minimum Distances (AMD) vs. XRD Similarity}
\label{sub:amd}
To assess how effectively diffraction-based similarity metrics capture underlying structural similarity, we compare each XRD similarity measure with the Average Minimum Distance (AMD) metric \cite{widdowson2022resolving}. AMD provides a continuous, geometry-based measure of similarity between periodic structures by computing the Earth Mover's Distance between the atomic pointwise distance distributions of the ground-truth and optimized structures. For each structure pair, we compare its AMD with the corresponding XRD similarity score and quantify their relationship using Mutual Information (MI), Pearson correlation, and Spearman correlation.

Ideally, higher XRD similarity should correspond to lower AMD, indicating that diffraction-space similarity aligns with structural similarity. MI captures overall statistical dependence between the two quantities, while Pearson and Spearman correlations describe linear and monotonic relationships, respectively.

As shown in Table~\ref{tab:mutual_info}, the relationship between structural and diffraction-based similarity is not uniform across metrics or noise levels. For lattice distortions, cosine similarity yields the highest mutual information (MI) at low noise (1.09), whereas entropy similarity performs best at moderate noise levels (0.53 and 0.21). At higher noise levels, all metrics exhibit similarly low MI, indicating a loss of structural correspondence. A comparable pattern is observed for coordinate distortions. In contrast, when symmetry constraints are enforced during optimization, MI increases substantially across all lattice noise levels, underscoring the benefit of restricting the search space to symmetry-consistent configurations.

The correlation results reported in Tables~\ref{tab:pearson_corr} and~\ref{tab:spear_corr} exhibit similar trends. For lattice distortions, no single XRD similarity metric consistently correlates with AMD in the unconstrained setting, whereas enforcing symmetry constraints improves both Pearson and Spearman correlations. For coordinate distortions, cosine and entropy similarity metrics show negative correlations with AMD, while the MSE-based metric yields a positive and relatively high Spearman correlation. This behavior can be attributed to the fact that MSE more strongly penalizes differences in peak intensities. Since low-level coordinate noise mainly affects peak heights through modifications to the structure factors rather than shifting peak positions, MSE captures these subtle structural perturbations more effectively.

\begin{figure}[ht]
    \centering
    % --- First row ---
    \begin{subfigure}[t]{0.4\textwidth}
        \centering
        \includegraphics[width=\textwidth]{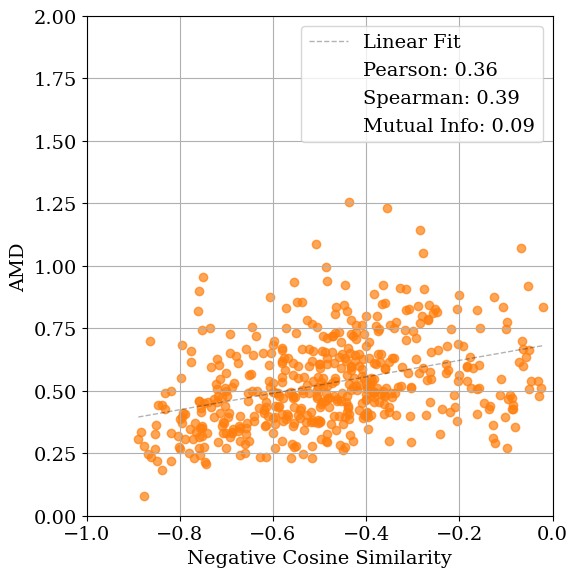}
        \caption{}
    \end{subfigure}
    \begin{subfigure}[t]{0.4\textwidth}
        \centering
        \includegraphics[width=\textwidth]{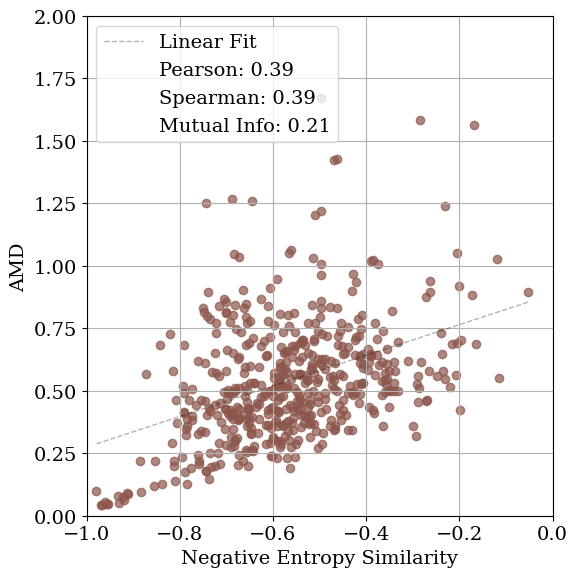}
        \caption{}
    \end{subfigure}
    
    % --- Second row ---
    \begin{subfigure}[t]{0.4\textwidth}
        \centering
        \includegraphics[width=\textwidth]{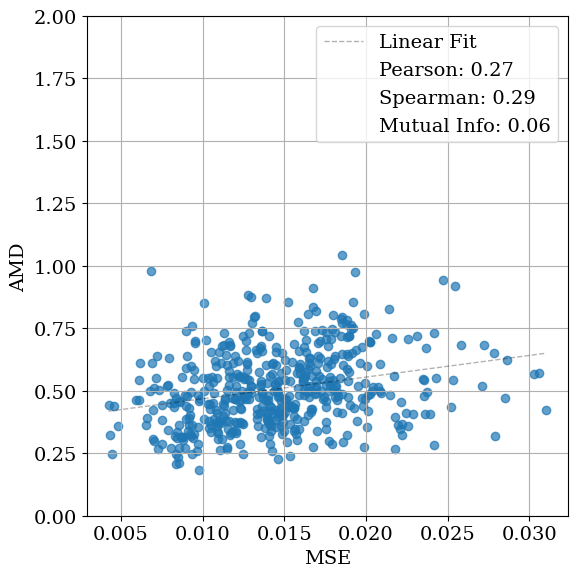}
        \caption{}
    \end{subfigure}
    \begin{subfigure}[t]{0.4\textwidth}
        \centering
        \includegraphics[width=\textwidth]{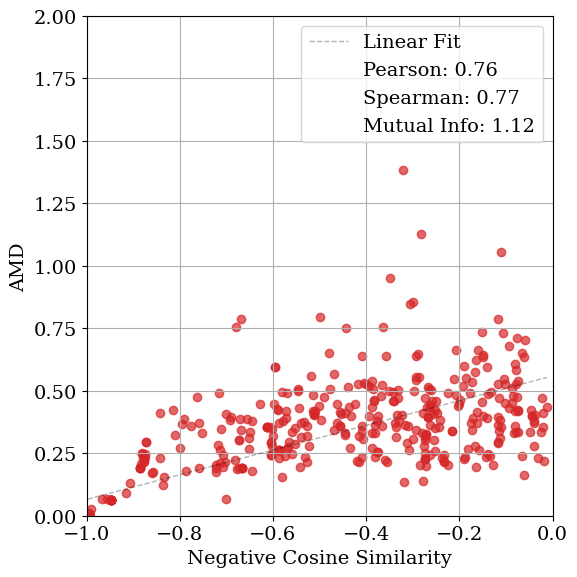}
        \caption{}
    \phantomcaption
    \end{subfigure}

    \caption{\textbf{Average Minimum Distance (AMD) vs. XRD Similarity Metrics.} (a) Cosine similarity. (b) Entropy similarity. (c) Mean squared error (MSE). (d) Cosine similarity with symmetry constraints applied during optimization. All panels compare structures obtained from XRD-based optimization following lattice distortions of 0.1. In this case, using entropy similarity as the optimization objective yields higher mutual information (MI) than cosine similarity or MSE. However, this trend is not consistent across all noise types and levels (see Table~\ref{tab:mutual_info}). Applying symmetry constraints improves MI as well as linear and Spearman correlations.}
    \label{fig:amd}
\end{figure}

Figure~\ref{fig:amd} visualizes these relationships. MSE exhibits the weakest correlation with AMD (low Pearson and Spearman coefficients), while cosine and entropy similarities show comparable correlations. Entropy similarity achieves slightly higher MI (0.20 vs. 0.09 for cosine), indicating a modestly stronger dependence between diffraction and geometric similarity under these conditions. Nevertheless, the data remain broadly scattered, revealing that all tested similarity metrics struggle to consistently distinguish structurally distinct configurations. When symmetry constraints are enforced (Fig.~\ref{fig:amd}d), the correspondence between AMD and cosine similarity improves substantially, both in MI (1.12) and correlation coefficients.

\begin{table}[ht]
\centering
\caption{Mutual Information (MI) Between Average Minimum Distances (AMD) and XRD Similarity Metrics for the Different Noise Types and Levels.}
\label{tab:mutual_info}
\begin{tabular}{c c c c c c}
\hline
\textbf{Noise Type} & \textbf{Noise Level} &  \makecell{\textbf{Cosine} \\ \textbf{Similarity}} & \textbf{MSE} & \makecell{\textbf{Entropy} \\ \textbf{Similarity}} & \makecell{\textbf{Cosine Similarity} \\ \textbf{+ Constraints}} \\
\hline
\multirow{5}{*}{Lattice} 
 & 0.01 & 1.09 & 0.25 & 0.76 & \textbf{1.40}\\
 & 0.05 & 0.12 & 0.16 & 0.53 & \textbf{1.55} \\
 & 0.1  & 0.09 & 0.06 & 0.21 & \textbf{1.12} \\
 & 0.15 & 0.03 & 0.02 & 0.02 & \textbf{0.9}  \\
 & 0.2  & 0.07 & 0.02 & 0.01 & \textbf{0.85} \\
\hline
\multirow{5}{*}{Coordinates} 
 & 0.001 & \textbf{1.05} & 0.32 & 0.68 & -\\
 & 0.005 & \textbf{0.83} & 0.48 & 0.53 & -\\
 & 0.01  & 0.44 & \textbf{0.67} & 0.56 & -\\
 & 0.05  & 0.96 & 0.36 & \textbf{1.41} & -\\
 & 0.1   & 0.77 & 0.32 & \textbf{1.06} & -\\
\hline
\end{tabular}
\end{table}

\begin{figure}[ht]
\centering
\includegraphics[width=0.99\textwidth] {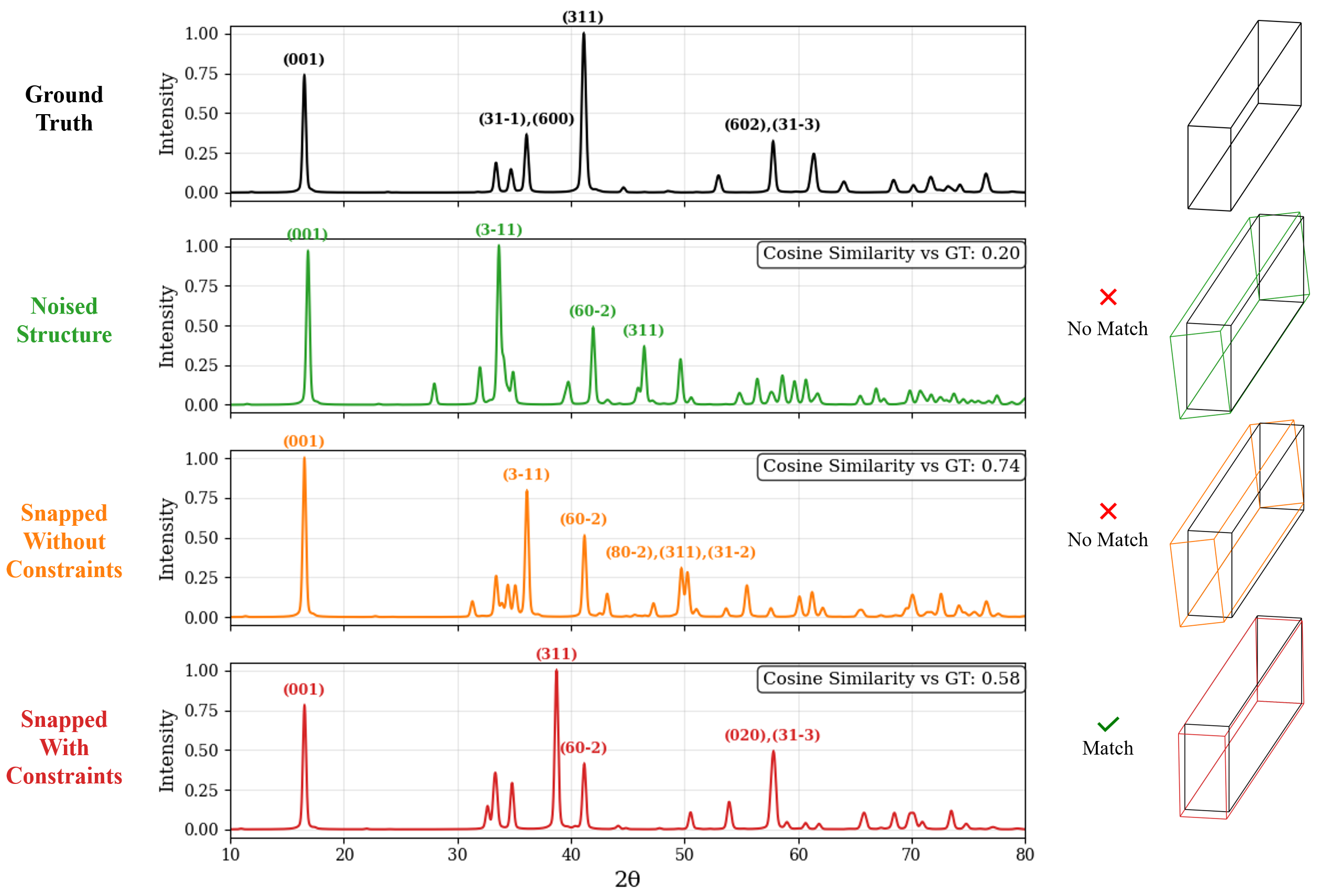}
\caption{\textbf{Lattice and XRD patterns of \ce{Na3MnCoNiO6}.} Each row shows the unit cell relative to the ground truth and corresponding XRD pattern. From top to bottom: ground truth; distorted lattice structure with 0.1 noise level; result of XRD-based GD optimization without constraints; and result of XRD-based GD optimization with symmetry-based constraints. For each, the cosine similarity to the ground truth pattern and the structure match status according to \texttt{StructureMatcher} are reported.}
\label{fig:xrd_agr}
\end{figure}

\subsection{On the Metrics Used for Measuring XRD Similarity}
Figure~\ref{fig:xrd_agr} illustrates structure optimization results, with and without lattice constraints. After applying noise, the structure no longer matches the ground truth according to the \texttt{StructureMatcher} metric, with the diffraction pattern exhibiting peak shifts and new reflections. The unconstrained GD optimizer converges to a local minimum where a few peaks align, resulting in a significant increase in cosine similarity (from 0.2 to 0.74). When crystal-family constraints are imposed, the optimizer recovers a structure that matches the ground truth within \texttt{StructureMatcher} tolerance, and the cosine similarity between the patterns increases (from 0.2 to 0.57). However, even in the constrained solution, aligned peaks do not necessarily correspond to identical crystallographic planes, labeled by the Miller indices $hkl$. E.g., the ground truth $311$ plane (top panel) appears slightly shifted in the constrained solution (bottom panel) but aligns with $60\bar{2}$, thereby contributing to the similarity score. This discrepancy in interatomic distances seems to remain within the tolerance.

\begin{figure}[ht]
\centering
\includegraphics[width=0.99\textwidth] {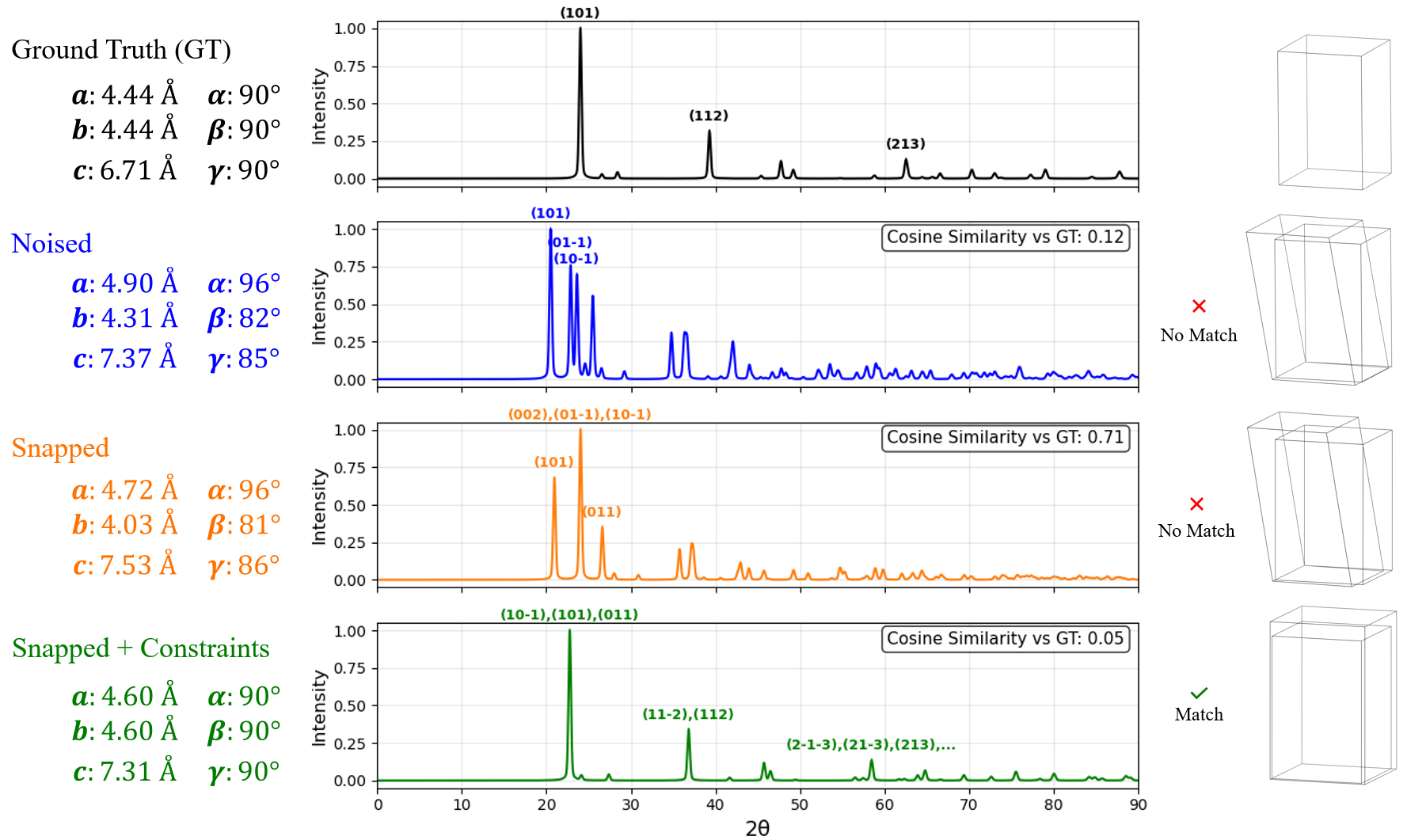}
\caption{\textbf{Lattice and XRD patterns of \ce{BPO4}.} Each row shows the lattice parameters, corresponding XRD pattern, and unit cell relative to the ground truth. From top to bottom: ground truth; distorted lattice structure with 0.1 noise level; result of XRD-based GD optimization without constraints; and result of XRD-based GD optimization with symmetry-based constraints. For each case, the cosine similarity to the ground truth pattern and the structure match status are reported.}
\label{fig:appnds_xrd_agr}
\end{figure}

In Figure~\ref{fig:appnds_xrd_agr}, the unconstrained GD optimizer converges to a distinct structure whose peaks overlap with those of the ground truth, achieving a deceptively high cosine similarity (0.71). In contrast, the symmetry-constrained optimization successfully recovers a matched structure, but the resulting diffraction pattern shows slightly shifted peaks, leading to a much lower cosine similarity (0.05). The convergence to this shallow minimum is likely driven by fluctuations along the symmetry axis, as elaborated in subsection~\ref{subsec:res_sym_const}. 

This counterintuitive outcome highlights a key limitation of using XRD pattern similarity metrics such as cosine similarity, MSE, and entropy similarity as the sole reconstruction objective: similar structures can yield dissimilar patterns, and conversely, distinct structures may appear similar, under such metrics. Because these metrics are insensitive to whether aligned peaks arise from the same atomic geometry, distinct structures can yield high similarity scores, while small geometric deviations can produce low ones. This results in a highly non-convex optimization landscape prone to spurious minima.

This limitation suggests that purely signal-based metrics are unlikely to yield a well-behaved (e.g., convex) optimization landscape. In the absence of smoother metrics, enforcing symmetry-based constraints (see~\ref{subsec:method_sym_const}) on the optimizer remains an effective way to navigate otherwise rugged landscapes.
    
\subsection{Comparison to Energy Relaxation}

\begin{figure}[ht]
    \centering
    % --- First row ---
    \begin{subfigure}[t]{0.48\textwidth}
        \centering
        \includegraphics[width=\textwidth]{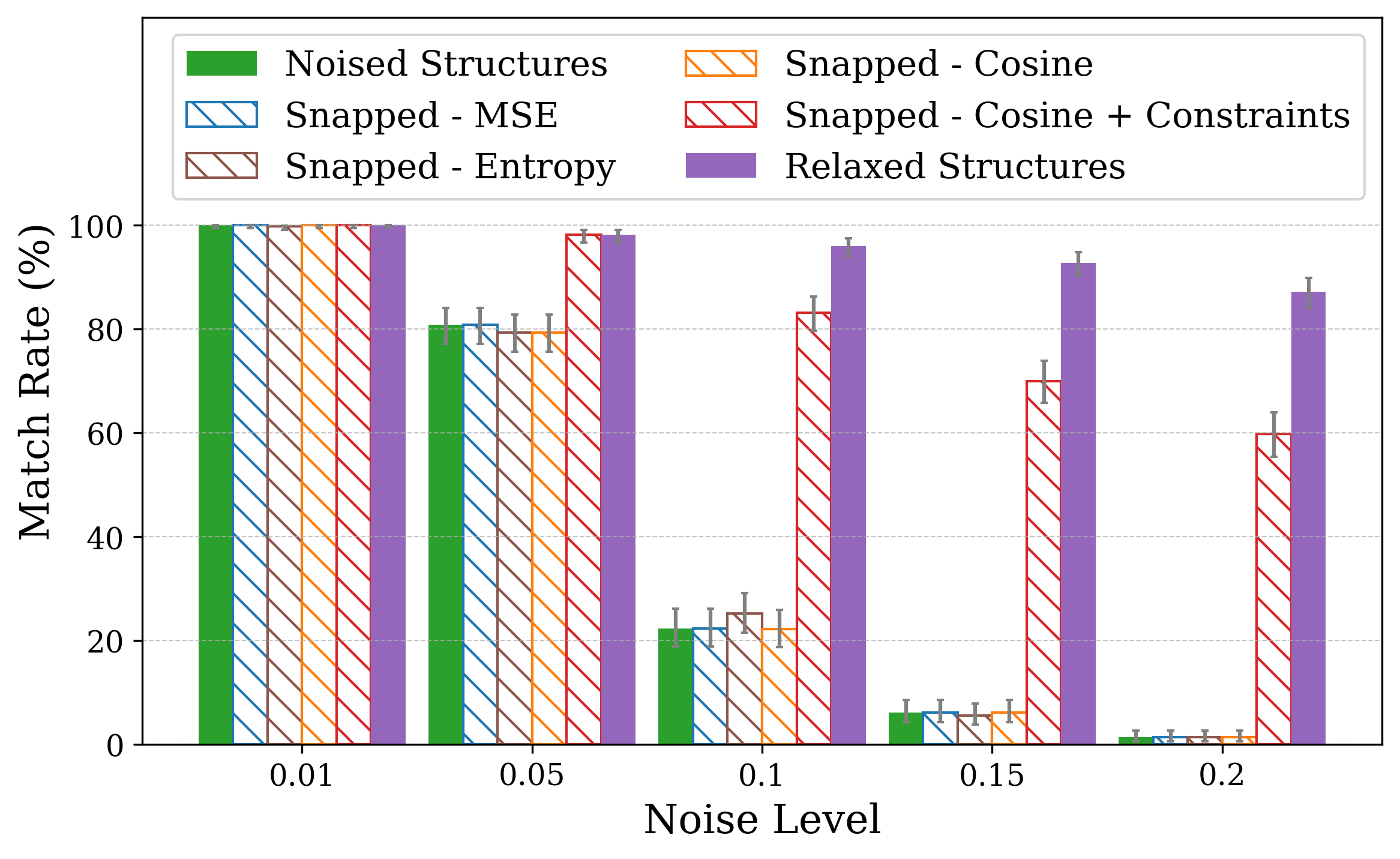}
        \caption*{Lattice noise}
    \end{subfigure}
    \hfill
    \begin{subfigure}[t]{0.48\textwidth}
        \centering
        \includegraphics[width=\textwidth]{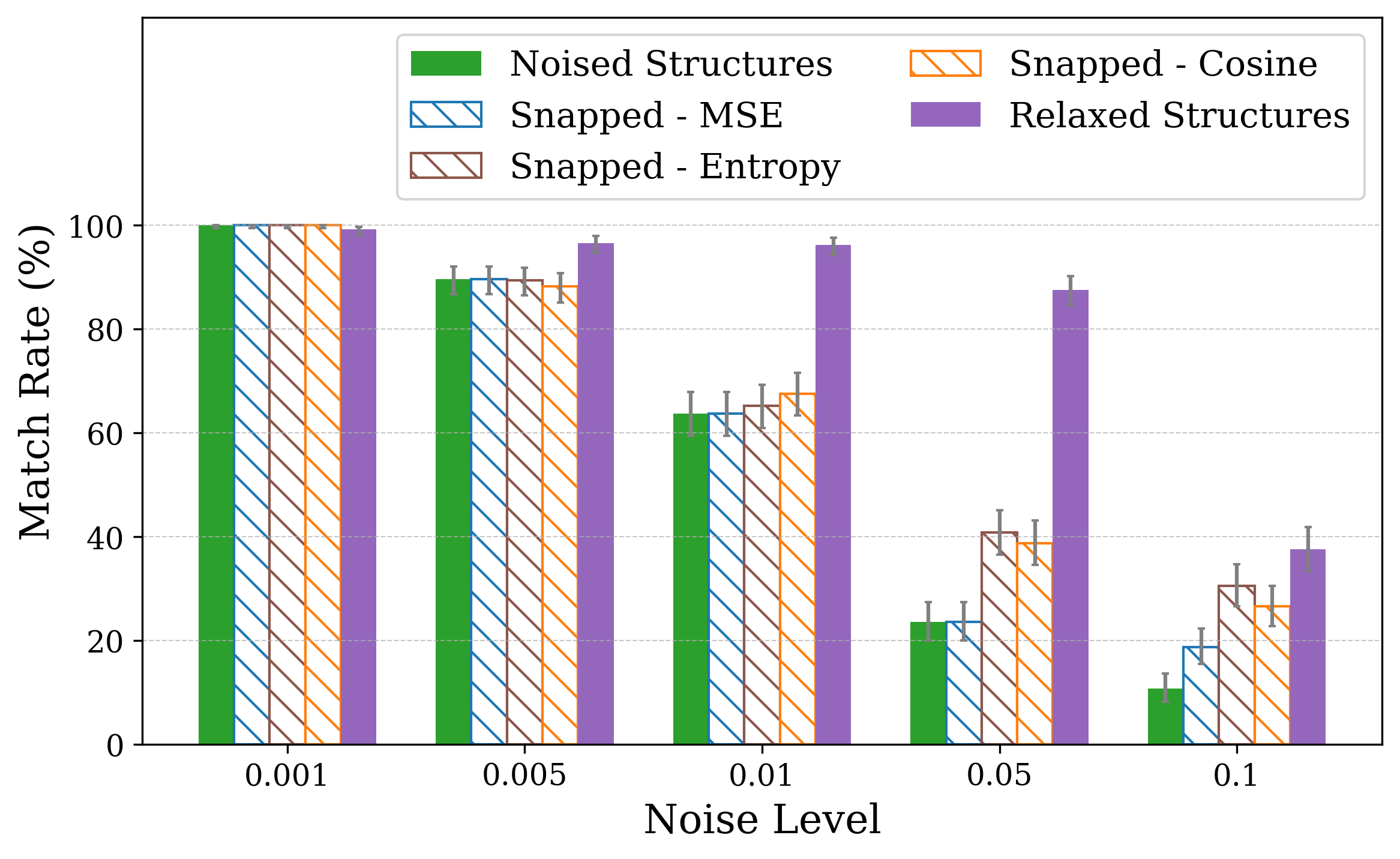}
        \caption*{Coordinate noise}
    \phantomcaption
    \label{subfig:match_comparison_energy}
    \end{subfigure}

    \vspace{0.5em} % Small space between rows

    % --- Second row ---
    \begin{subfigure}[t]{0.48\textwidth}
        \centering
        \includegraphics[width=\textwidth]{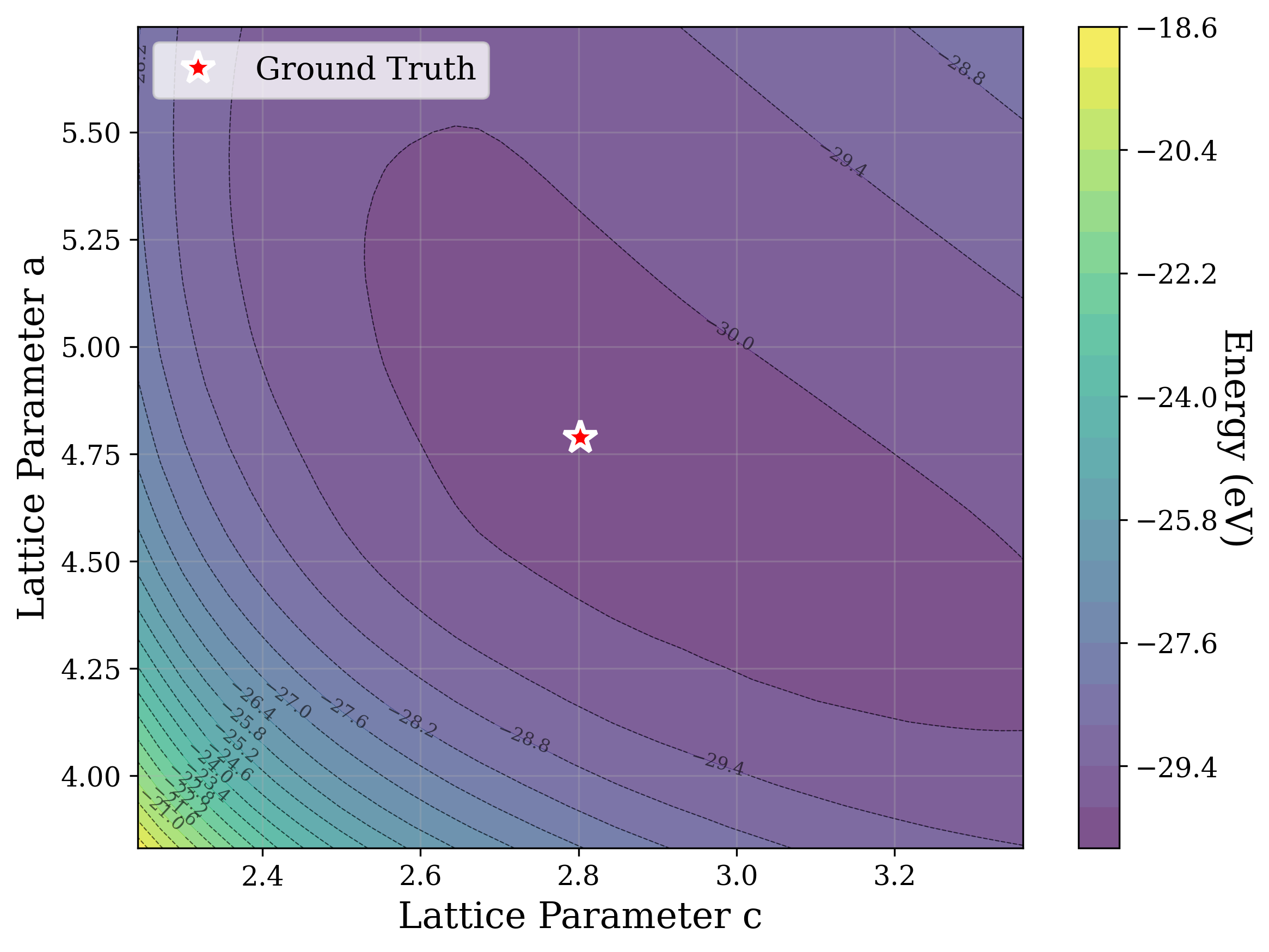}
        \caption*{\textit{$a,\ c$}}
    \end{subfigure}
    \hfill
    \begin{subfigure}[t]{0.48\textwidth}
        \centering
        \includegraphics[width=\textwidth]{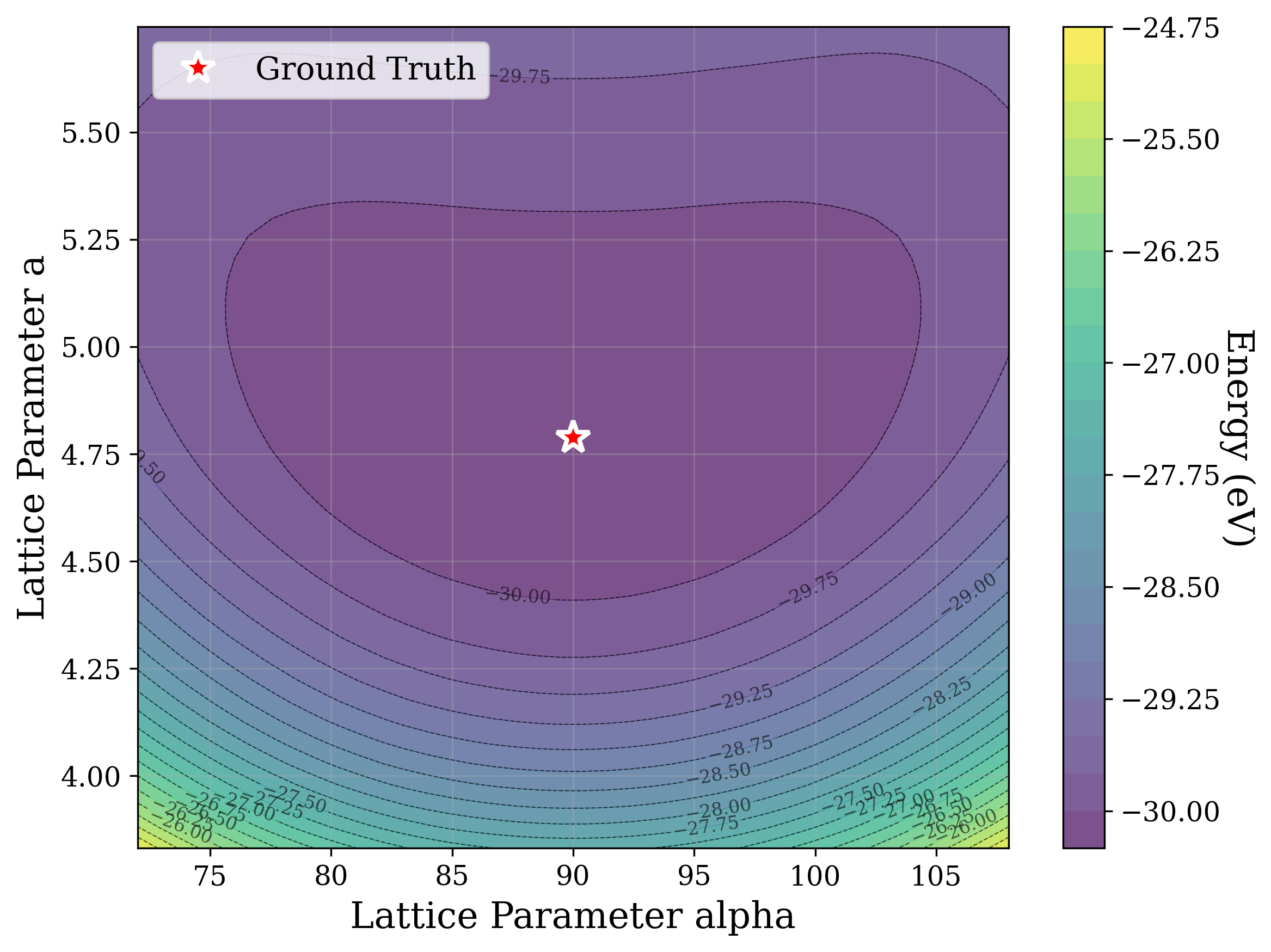}
        \caption*{\textit{$a,\ \alpha$}}
    \phantomcaption
    \label{subfig:energy_opt}
    \end{subfigure}

    \caption{\textbf{Comparing XRD-based optimization with energy relaxation}. \textbf{(Top)} Match rates from \texttt{StructureMatcher} with ($\text{ltol}=0.1, \ \text{stol}=0.2, \ \text{angle\_{tol}}=5^\circ$) under random lattice and coordinate perturbations. Snapped bars are the same as presented in Figure~\ref{fig:match_comparison}.
    Energy-based optimization consistently recovers the correct phase, except for high levels of coordinate noise, whereas XRD-based optimization struggles.
    \textbf{(Bottom)} 2D landscape of CHGNet \cite{deng2023chgnet} predicted energy as a function of lattice parameters of \ce{U2Ti}. Along lattice vectors, the energy landscape is smooth and locally convex.}
    \label{fig:combined_comparison_energy}
\end{figure}

Structural relaxation through potential energy minimization is often used \cite{deng2023chgnet, batatia2023foundation, chen2022universal, musaelian2023learning, wood2025family} to refine candidate structures. Figure~\ref{subfig:match_comparison_energy} shows that the universal ML interatomic potential (MLIP) CHGNet \cite{deng2023chgnet} accurately recovers structures matching the ground truth from the distorted state alone, except for a high level (0.1) of coordinate noise (Figure~\ref{subfig:energy_opt}). In that case, it is plausible that these larger distortions displace the system into the basin of attraction of a different local minimum on the potential energy surface. 

\begin{figure}[ht]
\centering
\setlength{\tabcolsep}{10pt} % adjust spacing between columns
\begin{tabular}{cc}
\includegraphics[width=0.44\textwidth]{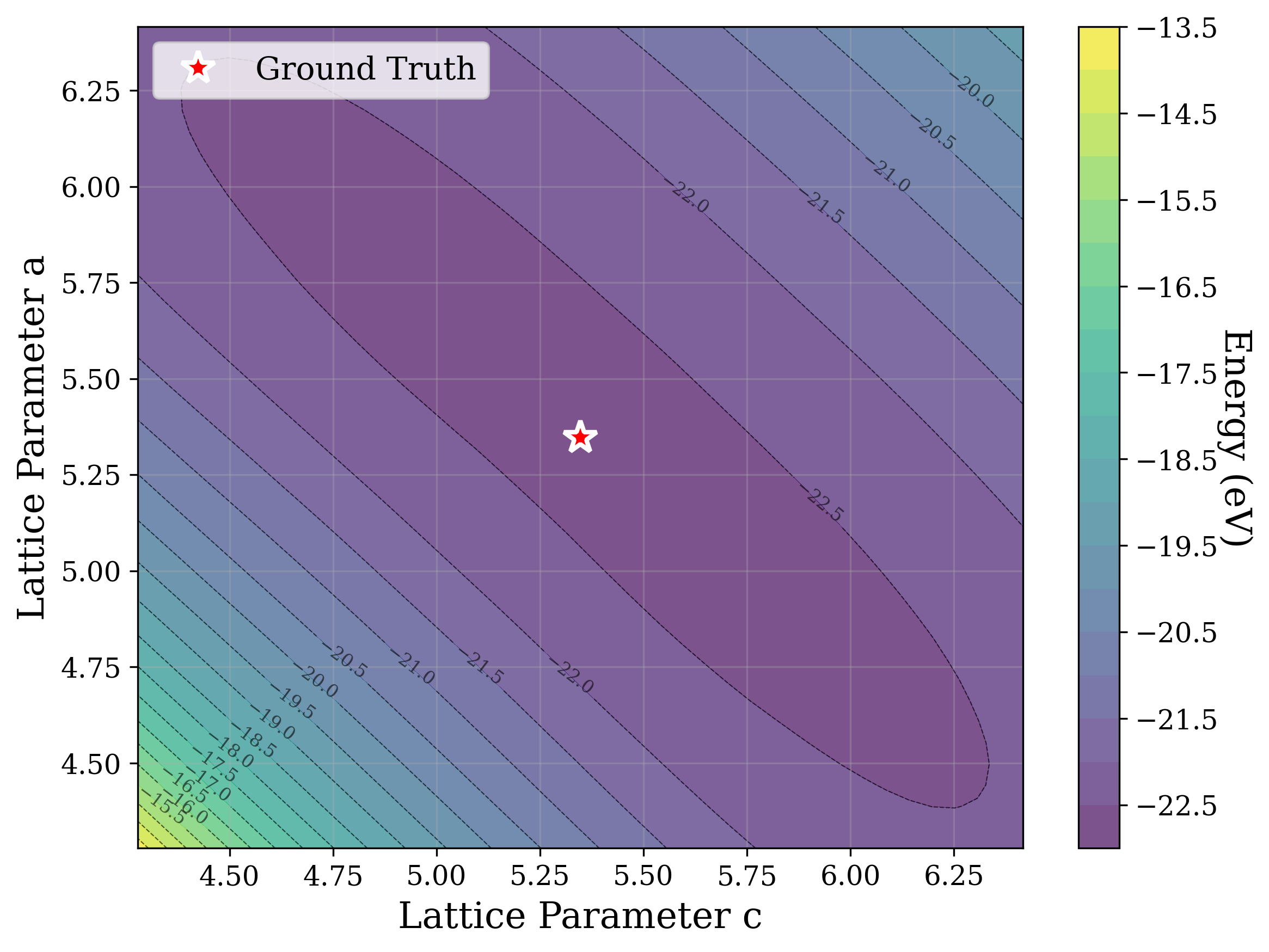} &
\includegraphics[width=0.44\textwidth]{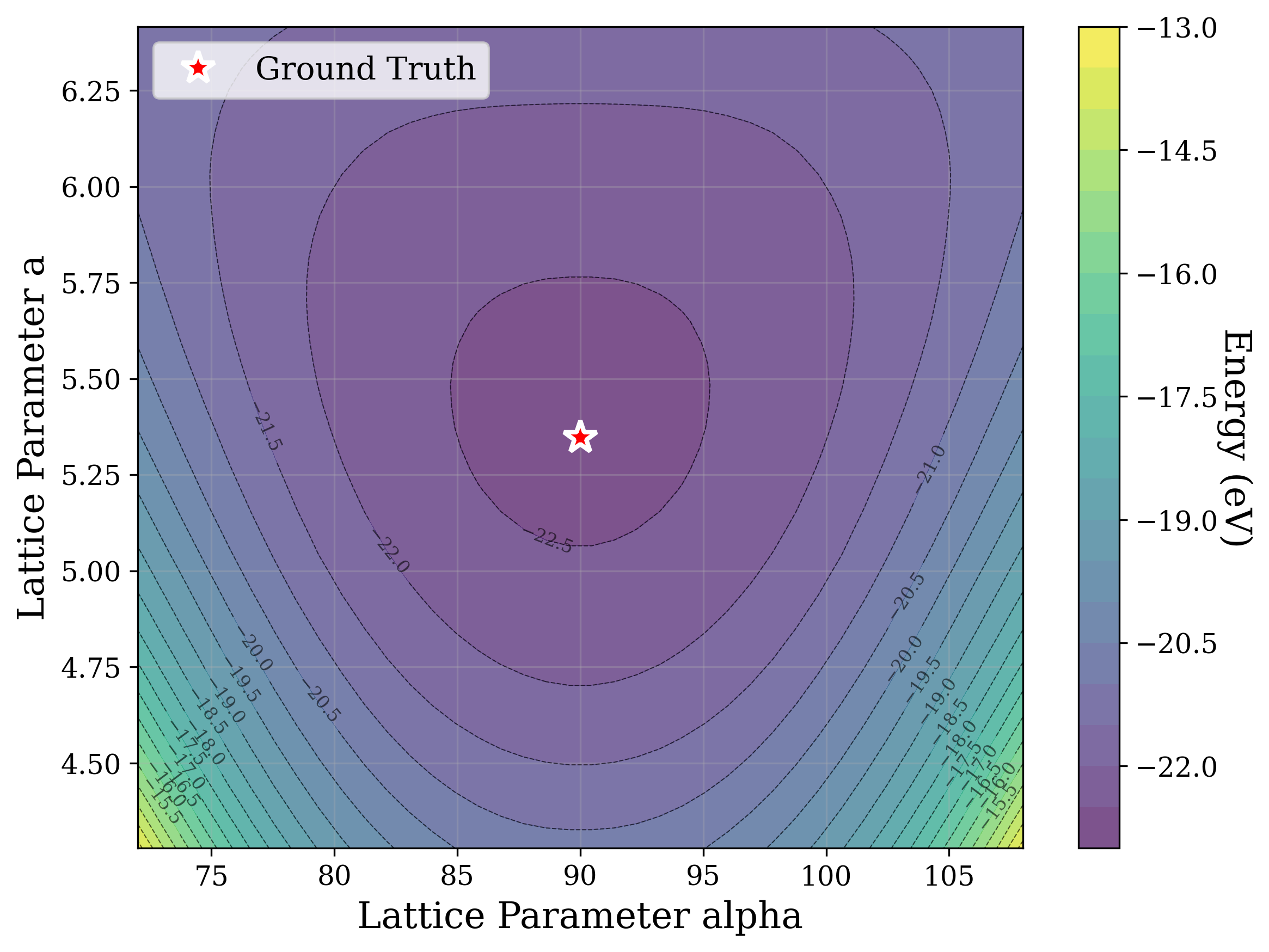} \\
\textit{$a, c$} & \textit{$a, \alpha$} \\[0.8em]

\includegraphics[width=0.44\textwidth]{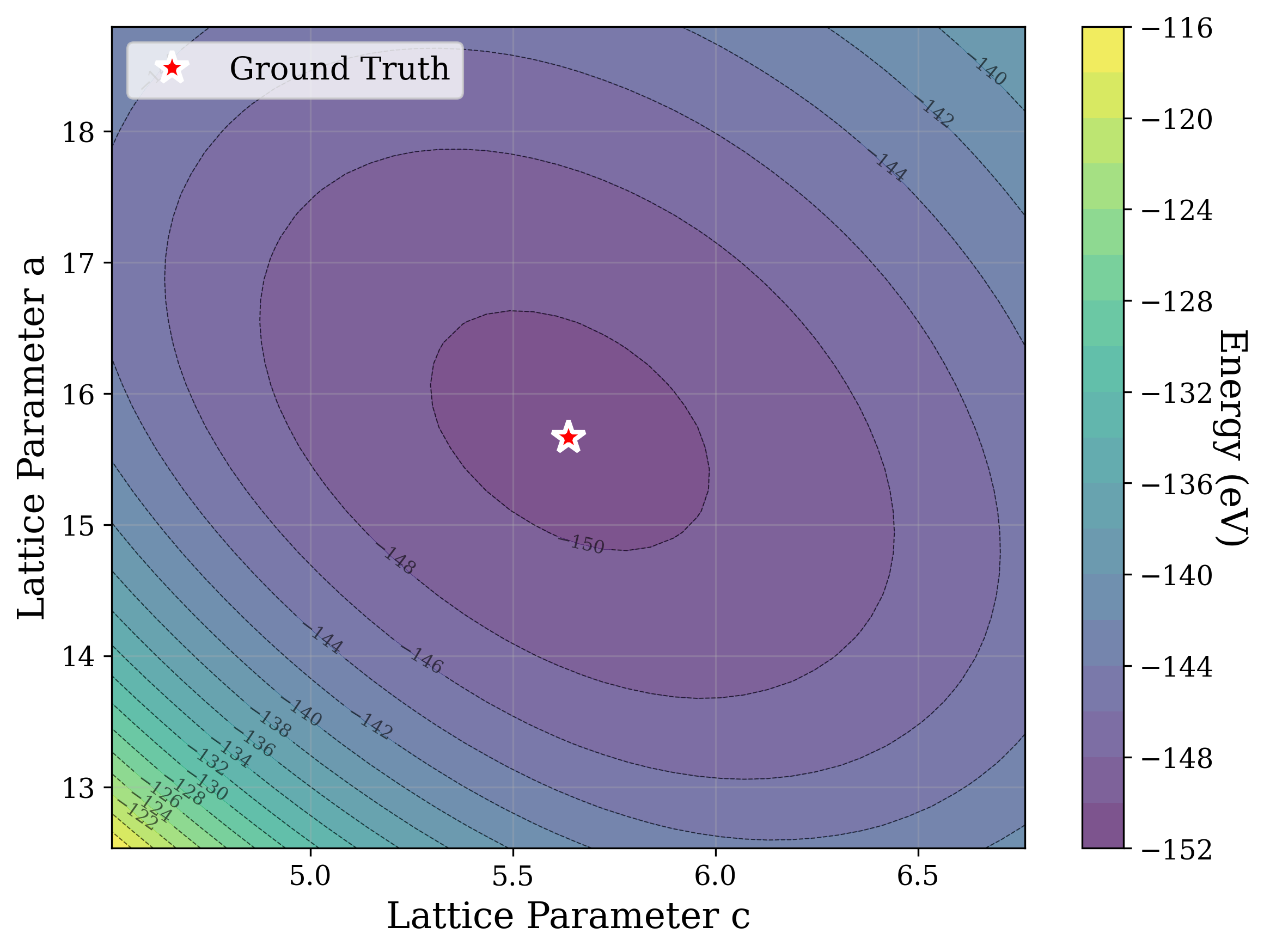} &
\includegraphics[width=0.44\textwidth]{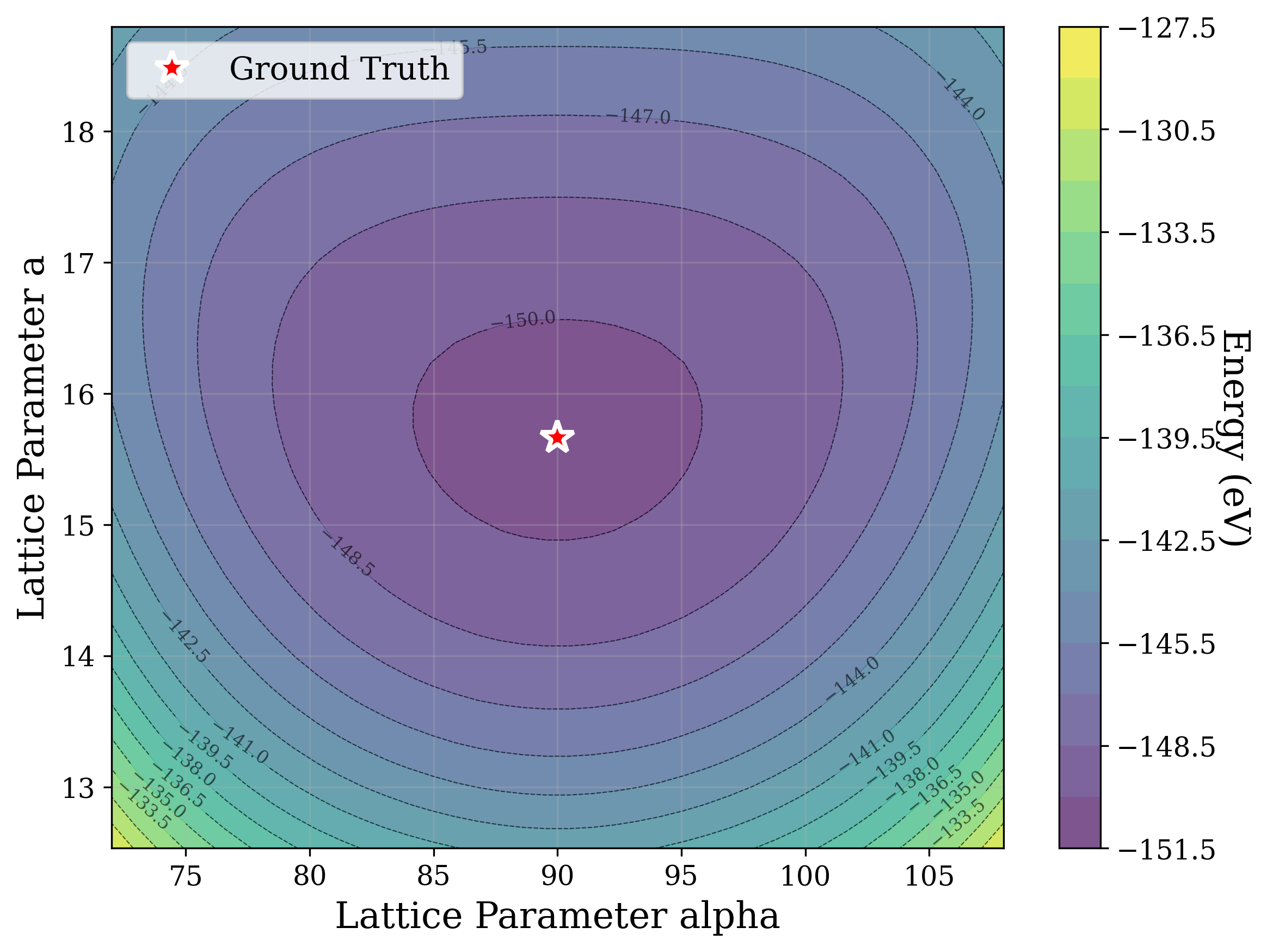} \\
\textit{$a, c$} & \textit{$a, \alpha$} \\[0.8em]

\includegraphics[width=0.44\textwidth]{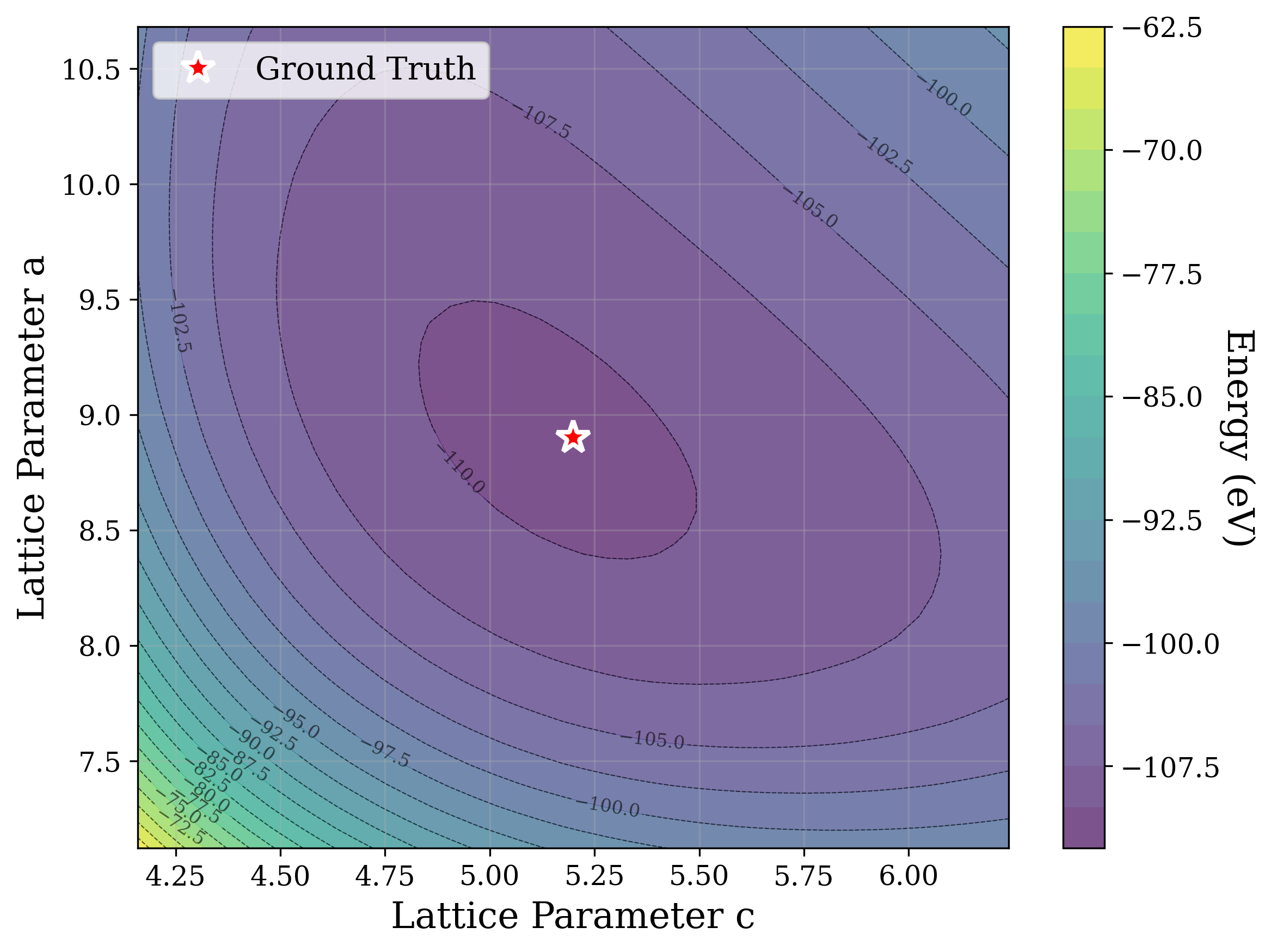} &
\includegraphics[width=0.44\textwidth]{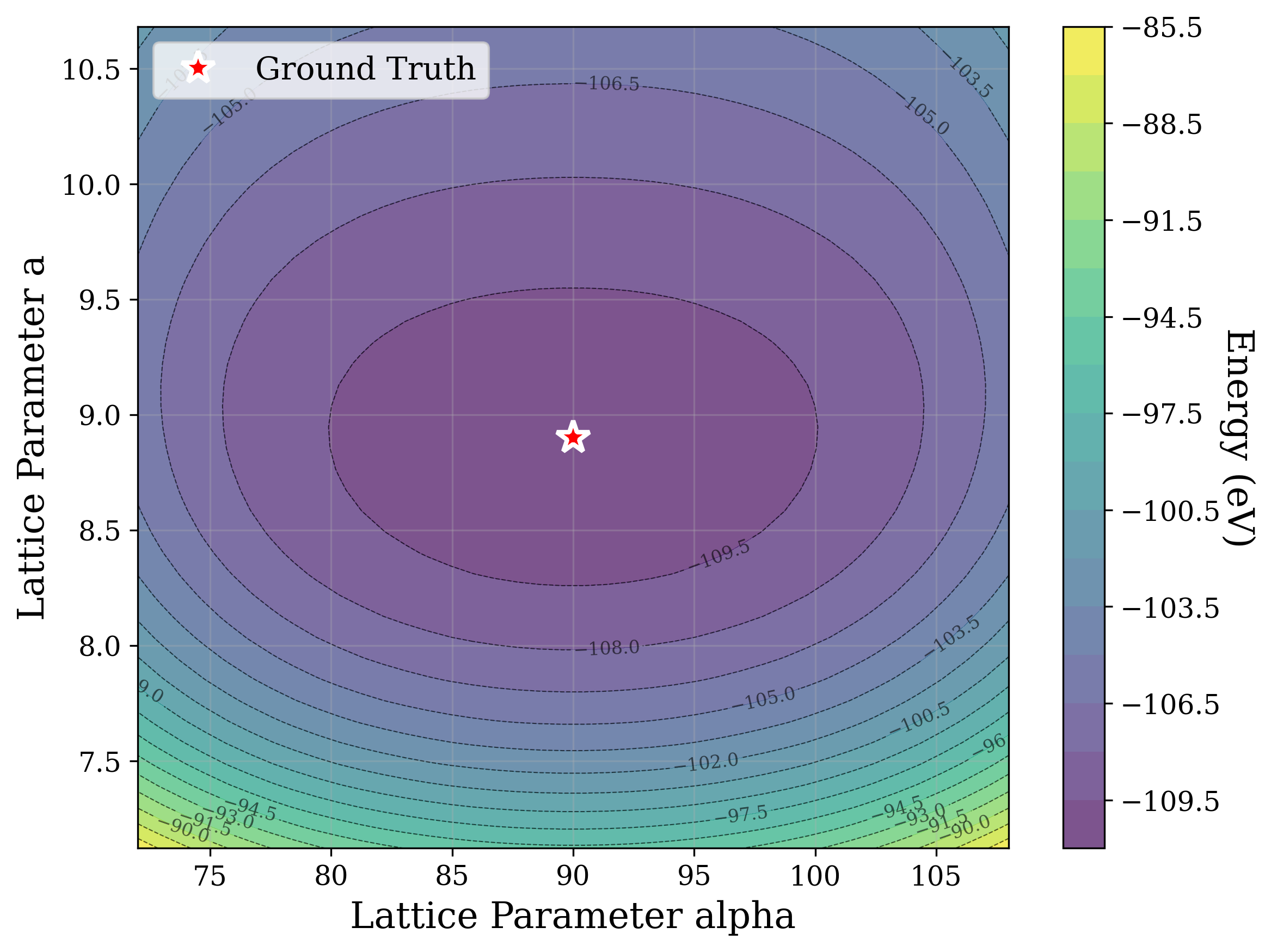} \\
\textit{$a, c$} & \textit{$a, \alpha$} \\[0.8em]
\end{tabular}

\caption{\textbf{2D landscape slices of Energy Relaxation for different structures from our study.}
(Top) Cubic \ce{Au2S}. (Middle) Monoclinic \ce{Na3MnCoNiO6}. (Bottom) Tetragonal \ce{Nd(Al2Cu)4}. Each slice is labeled with the distorted lattice parameters.}
\label{apndx:energy_slices}
\end{figure}

These results suggest that the loss landscape of energy relaxation is much smoother than that of XRD similarity \cite{deng2024overcoming}. Cross-sections of the energy optimization landscape are shown in Figures~\ref{subfig:energy_opt} and~\ref{apndx:energy_slices}, showcasing a smooth and convex behavior.

Energy relaxation and XRD-based optimization provide complementary signals: relaxation drives structures toward physically stable configurations, while XRD-based optimization attempts to match an observed pattern. Possible directions for future work can therefore include multi-objective optimization, which can combine the smoothness of the energy landscape with the structure-specific information captured by diffraction.

% While MLIPs accurately reconstruct structures under small, controlled deformations, their optimization landscape is not globally smooth. In the context of inverse XRD, the goal is to recover a specific structure from its diffraction pattern, corresponding to a particular minimum on the potential energy surface. Large distortions can displace the system into the basin of attraction of a different local minimum, as evident for coordinate perturbations of 10\% (Figure~\ref{subfig:energy_opt}). Moreover, strain-engineered structures \textemdash configurations in which external stress is used to stabilize properties unattainable at equilibrium \cite{hazra2024colossal, petrie2016enhancing} \textemdash cannot be recovered by MLIP relaxation. 
% XRD encodes structural information that is inaccessible to relaxation-based methods. Although MLIP relaxation succeeds under the limited, controlled distortions in this study, it cannot reliably recover the correct phase when the initial structure deviates substantially from it, or when the target configuration is not an energy minimum.

\section{Conclusion}
XRD provides a direct experimental link for generative crystal modeling, enabling the identification of novel phases. Our results highlight symmetry’s role in bridging XRD and structure, but also reveal that in some cases the XRD-to-structure landscape may remain non-convex even along symmetry axes, making post-hoc optimization difficult. We illustrate this with physically motivated random distortions, though generative models may introduce more complex biases. The observations in this work suggest that progress in inverse XRD can be made using new generative architectures that condition on XRD and embed symmetry as an inductive bias, with final refinements guided by energy relaxation.
%%%%%%%%%%%%%%%%%%%%%%%%%%%%%%%%%%%%%%%%%%%%%%%%%%
\clearpage
\bibliography{sn-bibliography}% common bib file

@article{riesel2024crystal,
  title={Crystal structure determination from powder diffraction patterns with generative machine learning},
  author={Riesel, Eric A and Mackey, Tsach and Nilforoshan, Hamed and Xu, Minkai and Badding, Catherine K and Altman, Alison B and Leskovec, Jure and Freedman, Danna E},
  journal={Journal of the American Chemical Society},
  volume={146},
  number={44},
  pages={30340--30348},
  year={2024},
  publisher={ACS Publications}
}

@article{ong2013python,
  title={Python Materials Genomics (pymatgen): A robust, open-source python library for materials analysis},
  author={Ong, Shyue Ping and Richards, William Davidson and Jain, Anubhav and Hautier, Geoffroy and Kocher, Michael and Cholia, Shreyas and Gunter, Dan and Chevrier, Vincent L and Persson, Kristin A and Ceder, Gerbrand},
  journal={Computational Materials Science},
  volume={68},
  pages={314--319},
  year={2013},
  publisher={Elsevier}
}

@article{cannelli2022atomic,
  title={Atomic-level description of thermal fluctuations in inorganic lead halide perovskites},
  author={Cannelli, Oliviero and Wiktor, Julia and Colonna, Nicola and Leroy, Ludmila and Puppin, Michele and Bacellar, Camila and Sadykov, Ilia and Krieg, Franziska and Smolentsev, Grigory and Kovalenko, Maksym V and others},
  journal={The Journal of Physical Chemistry Letters},
  volume={13},
  number={15},
  pages={3382--3391},
  year={2022},
  publisher={ACS Publications}
}

@article{brivio2015lattice,
  title={Lattice dynamics and vibrational spectra of the orthorhombic, tetragonal, and cubic phases of methylammonium lead iodide},
  author={Brivio, Federico and Frost, Jarvist M and Skelton, Jonathan M and Jackson, Adam J and Weber, Oliver J and Weller, Mark T and Goni, Alejandro R and Leguy, Aur{\'e}lien MA and Barnes, Piers RF and Walsh, Aron},
  journal={Physical Review B},
  volume={92},
  number={14},
  pages={144308},
  year={2015},
  publisher={APS}
}

@article{delgado2013effects,
  title={Effects of thermal expansion of the crystal lattice on x-ray crystal spectrometers used for fusion research},
  author={Delgado-Aparicio, L and Bitter, M and Podpaly, Y and Rice, J and Burke, W and Del Rio, M Sanchez and Beiersdorfer, P and Bell, R and Feder, R and Gao, C and others},
  journal={Plasma Physics and Controlled Fusion},
  volume={55},
  number={12},
  pages={125011},
  year={2013},
  publisher={IOP Publishing}
}

@article{guo2025ab,
  title={Ab initio structure solutions from nanocrystalline powder diffraction data via diffusion models},
  author={Guo, Gabe and Saidi, Tristan Luca and Terban, Maxwell W and Valsecchi, Michele and Billinge, Simon JL and Lipson, Hod},
  journal={Nature Materials},
  pages={1--9},
  year={2025},
  publisher={Nature Publishing Group}
}

@article{guo2024towards,
  title={Towards end-to-end structure determination from x-ray diffraction data using deep learning},
  author={Guo, Gabe and Goldfeder, Judah and Lan, Ling and Ray, Aniv and Yang, Albert Hanming and Chen, Boyuan and Billinge, Simon JL and Lipson, Hod},
  journal={npj Computational Materials},
  volume={10},
  number={1},
  pages={209},
  year={2024},
  publisher={Nature Publishing Group UK London}
}

@article{levy2025symmcd,
  title={SymmCD: Symmetry-Preserving crystal generation with diffusion models},
  author={Levy, Daniel and Panigrahi, Siba Smarak and Kaba, S{\'e}kou-Oumar and Zhu, Qiang and Lee, Kin Long Kelvin and Galkin, Mikhail and Miret, Santiago and Ravanbakhsh, Siamak},
  journal={arXiv preprint arXiv:2502.03638},
  year={2025}
}

@article{kazeev2025wyckoff,
  title={Wyckoff transformer: Generation of symmetric crystals},
  author={Kazeev, Nikita and Nong, Wei and Romanov, Ignat and Zhu, Ruiming and Ustyuzhanin, Andrey and Yamazaki, Shuya and Hippalgaonkar, Kedar},
  journal={arXiv preprint arXiv:2503.02407},
  year={2025}
}

@inproceedings{jiao2024space,
  title={Space Group Constrained Crystal Generation},
  author={Jiao, Rui and Huang, Wenbing and Liu, Yu and Zhao, Deli and Liu, Yang},
  booktitle={The Twelfth International Conference on Learning Representations},
  year={2024},
}

@inproceedings{xie2021crystal,
  title={Crystal Diffusion Variational Autoencoder for Periodic Material Generation},
  author={Xie, Tian and Fu, Xiang and Ganea, Octavian-Eugen and Barzilay, Regina and Jaakkola, Tommi S},
  booktitle={International Conference on Learning Representations},
  year={2021},
}

@article{hauptman1991phase,
  title={The phase problem of x-ray crystallography},
  author={Hauptman, Herbert A},
  journal={Reports on Progress in Physics},
  volume={54},
  number={11},
  pages={1427},
  year={1991},
  publisher={IOP Publishing}
}

@article{bragg1914analysis,
  title={The analysis of crystals by the X-ray spectrometer},
  author={Bragg, William Lawrence},
  journal={Proceedings of the Royal society of London. Series A, Containing papers of a mathematical and physical character},
  volume={89},
  number={613},
  pages={468--489},
  year={1914},
  publisher={The Royal Society London}
}

@book{hammond2015basics,
  title={The basics of crystallography and diffraction},
  author={Hammond, Christopher},
  volume={21},
  year={2015},
  publisher={Oxford university press}
}

@incollection{allen2006cambridge,
  title={The Cambridge Structural Database (CSD)},
  author={Allen, FH and Hoy, VJ},
  booktitle={International Tables for Crystallography Volume F: Crystallography of biological macromolecules},
  pages={663--668},
  year={2006},
  publisher={Springer}
}

@article{gates2019powder,
  title={The powder diffraction file: a quality materials characterization database},
  author={Gates-Rector, Stacy and Blanton, Thomas},
  journal={Powder diffraction},
  volume={34},
  number={4},
  pages={352--360},
  year={2019},
  publisher={Cambridge University Press}
}

@article{szymanski2024integrated,
  title={Integrated analysis of X-ray diffraction patterns and pair distribution functions for machine-learned phase identification},
  author={Szymanski, Nathan J and Fu, Sean and Persson, Ellen and Ceder, Gerbrand},
  journal={Npj computational materials},
  volume={10},
  number={1},
  pages={45},
  year={2024},
  publisher={Nature Publishing Group UK London}
}

@misc{holder2019tutorial,
  title={Tutorial on powder X-ray diffraction for characterizing nanoscale materials},
  author={Holder, Cameron F and Schaak, Raymond E},
  journal={ACS nano},
  volume={13},
  number={7},
  pages={7359--7365},
  year={2019},
  publisher={ACS Publications}
}

@article{chandra1999analysis,
  title={Analysis and characterization of data from twinned crystals},
  author={Chandra, N and Acharya, K Ravi and Moody, PCE},
  journal={Biological Crystallography},
  volume={55},
  number={10},
  pages={1750--1758},
  year={1999},
  publisher={International Union of Crystallography}
}

@article{rietveld1969profile,
  title={A profile refinement method for nuclear and magnetic structures},
  author={Rietveld, Hugo M},
  journal={Journal of applied Crystallography},
  volume={2},
  number={2},
  pages={65--71},
  year={1969},
  publisher={International Union of Crystallography}
}

@article{biwer2025spotlight,
  title={Spotlight: efficient automated global optimization in rietveld analysis of diffraction data},
  author={Biwer, CM and Feng, Z and Finstad, D and McDonnell, M and Knezevic, M and McKerns, M and Savage, DJ and Vogel, SC},
  journal={Scientific Reports},
  volume={15},
  number={1},
  pages={8358},
  year={2025},
  publisher={Nature Publishing Group UK London}
}

@article{deng2022preparation,
  title={Preparation of In/Sn nanoparticles (In3Sn and InSn4) by wet chemical one-step reduction and performance study},
  author={Deng, Huaming and Wang, Kaijun and Duan, Yunbiao and Zhang, Weijun and Hu, Jin},
  journal={Coatings},
  volume={12},
  number={4},
  pages={429},
  year={2022},
  publisher={MDPI}
}

@article{gu2017sustainable,
  title={A sustainable approach to separate and recover indium and tin from spent indium--tin oxide targets},
  author={Gu, Shuai and Fu, Bitian and Dodbiba, Gjergj and Fujita, Toyohisa and Fang, Baizeng},
  journal={RSC Advances},
  volume={7},
  number={82},
  pages={52017--52023},
  year={2017},
  publisher={Royal Society of Chemistry}
}

@article{wang2024comparison,
  title={Comparison of high-speed shear properties of low-temperature Sn-Bi/Cu and Sn-In/Cu solder joints},
  author={Wang, Qin and Cai, Shanshan and Yang, Shiyu and Yu, Yongjian and Wan, Yongkang and Peng, Jubo and Wang, Jiajun and Wang, Xiaojing},
  journal={Journal of Materials Science: Materials in Electronics},
  volume={35},
  number={8},
  pages={576},
  year={2024},
  publisher={Springer}
}

@article{el2023robust,
  title={Robust effects of In, Fe, and Co additions on microstructures, thermal, and mechanical properties of hypoeutectic Sn--Zn-based lead-free solder alloy},
  author={El-Taher, AM and Mansour, SA and Lotfy, IH},
  journal={Journal of Materials Science: Materials in Electronics},
  volume={34},
  number={7},
  pages={599},
  year={2023},
  publisher={Springer}
}

@article{ayeshamariam2014morphological,
  title={Morphological, structural, and gas-sensing characterization of tin-doped indium oxide nanoparticles},
  author={Ayeshamariam, A and Kashif, M and Bououdina, M and Hashim, U and Jayachandran, M and Ali, ME},
  journal={Ceramics International},
  volume={40},
  number={1},
  pages={1321--1328},
  year={2014},
  publisher={Elsevier}
}

@article{ashokkumar2024green,
  title={Green synthesis of silver and copper nanoparticles and their composites using Ocimum sanctum leaf extract displayed enhanced antibacterial, antioxidant and anticancer potentials},
  author={Ashokkumar, M and Palanisamy, K and Ganesh Kumar, A and Muthusamy, C and Senthil Kumar, KJ},
  journal={Artificial Cells, Nanomedicine, and Biotechnology},
  volume={52},
  number={1},
  pages={438--448},
  year={2024},
  publisher={Taylor \& Francis}
}

@article{jain2013commentary,
  title={Commentary: The Materials Project: A materials genome approach to accelerating materials innovation},
  author={Jain, Anubhav and Ong, Shyue Ping and Hautier, Geoffroy and Chen, Wei and Richards, William Davidson and Dacek, Stephen and Cholia, Shreyas and Gunter, Dan and Skinner, David and Ceder, Gerbrand and others},
  journal={APL materials},
  volume={1},
  number={1},
  year={2013},
  publisher={AIP Publishing}
}

@article{deng2023chgnet,
  title={CHGNet as a pretrained universal neural network potential for charge-informed atomistic modelling},
  author={Deng, Bowen and Zhong, Peichen and Jun, KyuJung and Riebesell, Janosh and Han, Kevin and Bartel, Christopher J and Ceder, Gerbrand},
  journal={Nature Machine Intelligence},
  volume={5},
  number={9},
  pages={1031--1041},
  year={2023},
  publisher={Nature Publishing Group UK London}
}

@article{chen2022universal,
  title={A universal graph deep learning interatomic potential for the periodic table},
  author={Chen, Chi and Ong, Shyue Ping},
  journal={Nature Computational Science},
  volume={2},
  number={11},
  pages={718--728},
  year={2022},
  publisher={Nature Publishing Group US New York}
}

@article{batatia2023foundation,
  title={A foundation model for atomistic materials chemistry},
  author={Batatia, Ilyes and Benner, Philipp and Chiang, Yuan and Elena, Alin M and Kov{\'a}cs, D{\'a}vid P and Riebesell, Janosh and Advincula, Xavier R and Asta, Mark and Avaylon, Matthew and Baldwin, William J and others},
  journal={arXiv preprint arXiv:2401.00096},
  year={2023}
}

@article{musaelian2023learning,
  title={Learning local equivariant representations for large-scale atomistic dynamics},
  author={Musaelian, Albert and Batzner, Simon and Johansson, Anders and Sun, Lixin and Owen, Cameron J and Kornbluth, Mordechai and Kozinsky, Boris},
  journal={Nature Communications},
  volume={14},
  number={1},
  pages={579},
  year={2023},
  publisher={Nature Publishing Group UK London}
}

@article{wood2025family,
  title={UMA: A Family of Universal Models for Atoms},
  author={Wood, Brandon M and Dzamba, Misko and Fu, Xiang and Gao, Meng and Shuaibi, Muhammed and Barroso-Luque, Luis and Abdelmaqsoud, Kareem and Gharakhanyan, Vahe and Kitchin, John R and Levine, Daniel S and others},
  journal={arXiv preprint arXiv:2506.23971},
  year={2025}
}

@article{parackal2024identifying,
  title={Identifying crystal structures beyond known prototypes from x-ray powder diffraction spectra},
  author={Parackal, Abhijith S and Goodall, Rhys EA and Faber, Felix A and Armiento, Rickard},
  journal={Physical Review Materials},
  volume={8},
  number={10},
  pages={103801},
  year={2024},
  publisher={APS}
}

@article{lee2023creation,
  title={Creation of crystal structure reproducing X-ray diffraction pattern without using database},
  author={Lee, Joohwi and Oba, Junpei and Ohba, Nobuko and Kajita, Seiji},
  journal={npj Computational Materials},
  volume={9},
  number={1},
  pages={135},
  year={2023},
  publisher={Nature Publishing Group UK London}
}

@article{johansen2025decifer,
  title={deCIFer: Crystal Structure Prediction from Powder Diffraction Data using Autoregressive Language Models},
  author={Johansen, Frederik Lizak and Friis-Jensen, Ulrik and Dam, Erik Bj{\o}rnager and Jensen, Kirsten Marie {\O}rnsbjerg and Mercado, Roc{\'\i}o and Selvan, Raghavendra},
  journal={arXiv preprint arXiv:2502.02189},
  year={2025}
}

@article{gevorkov2019xgandalf,
  title={XGANDALF--extended gradient descent algorithm for lattice finding},
  author={Gevorkov, Yaroslav and Yefanov, Oleksandr and Barty, Anton and White, Thomas A and Mariani, Valerio and Brehm, Wolfgang and Tolstikova, Aleksandra and Grigat, R-R and Chapman, Henry N},
  journal={Foundations of Crystallography},
  volume={75},
  number={5},
  pages={694--704},
  year={2019},
  publisher={International Union of Crystallography}
}

@article{otero2024powder,
  title={Powder-diffraction-based structural comparison for crystal structure prediction without prior indexing},
  author={Otero-de-la-Roza, Alberto},
  journal={Applied Crystallography},
  volume={57},
  number={5},
  year={2024},
  publisher={International Union of Crystallography}
}

@article{suzuki2020symmetry,
  title={Symmetry prediction and knowledge discovery from X-ray diffraction patterns using an interpretable machine learning approach},
  author={Suzuki, Yuta and Hino, Hideitsu and Hawai, Takafumi and Saito, Kotaro and Kotsugi, Masato and Ono, Kanta},
  journal={Scientific reports},
  volume={10},
  number={1},
  pages={21790},
  year={2020},
  publisher={Nature Publishing Group UK London}
}

@article{cao2024simxrd,
  title={SimXRD-4M: Big Simulated X-ray Diffraction Data Accelerate the Crystal Symmetry Classification},
  author={Cao, Bin and Liu, Yang and Zheng, Zinan and Tan, Ruifeng and Li, Jia and Zhang, Tong-yi},
  journal={arXiv preprint arXiv:2406.15469},
  year={2024}
}

@article{lee2022powder,
  title={Powder X-ray diffraction pattern is all you need for machine-learning-based symmetry identification and property prediction},
  author={Lee, Byung Do and Lee, Jin-Woong and Park, Woon Bae and Park, Joonseo and Cho, Min-Young and Pal Singh, Satendra and Pyo, Myoungho and Sohn, Kee-Sun},
  journal={Advanced Intelligent Systems},
  volume={4},
  number={7},
  pages={2200042},
  year={2022},
  publisher={Wiley Online Library}
}

@article{brown2001interval,
  title={Interval estimation for a binomial proportion},
  author={Brown, Lawrence D and Cai, T Tony and DasGupta, Anirban},
  journal={Statistical science},
  volume={16},
  number={2},
  pages={101--133},
  year={2001},
  publisher={Institute of Mathematical Statistics}
}

@article{qian1999momentum,
  title={On the momentum term in gradient descent learning algorithms},
  author={Qian, Ning},
  journal={Neural networks},
  volume={12},
  number={1},
  pages={145--151},
  year={1999},
  publisher={Elsevier}
}

@article{deng2024overcoming,
  title={Overcoming systematic softening in universal machine learning interatomic potentials by fine-tuning},
  author={Deng, Bowen and Choi, Yunyeong and Zhong, Peichen and Riebesell, Janosh and Anand, Shashwat and Li, Zhuohan and Jun, KyuJung and Persson, Kristin A and Ceder, Gerbrand},
  journal={arXiv preprint arXiv:2405.07105},
  year={2024}
}

@article{li2025powder,
  title={Powder diffraction crystal structure determination using generative models},
  author={Li, Qi and Jiao, Rui and Wu, Liming and Zhu, Tiannian and Huang, Wenbing and Jin, Shifeng and Liu, Yang and Weng, Hongming and Chen, Xiaolong},
  journal={Nature Communications},
  volume={16},
  number={1},
  pages={7428},
  year={2025},
  publisher={Nature Publishing Group UK London}
}

@article{racioppi2025powder,
  title={Powder X-ray diffraction assisted evolutionary algorithm for crystal structure prediction},
  author={Racioppi, Stefano and Otero-de-la-Roza, Alberto and Hajinazar, Samad and Zurek, Eva},
  journal={Digital Discovery},
  volume={4},
  number={1},
  pages={73--83},
  year={2025},
  publisher={Royal Society of Chemistry}
}

@article{hernandez2017using,
  title={Using similarity metrics to quantify differences in high-throughput data sets: application to X-ray diffraction patterns},
  author={Hern{\'a}ndez-Rivera, Efra{\'\i}n and Coleman, Shawn P and Tschopp, Mark A},
  journal={ACS combinatorial science},
  volume={19},
  number={1},
  pages={25--36},
  year={2017},
  publisher={ACS Publications}
}

@article{li2021spectral,
  title={Spectral entropy outperforms MS/MS dot product similarity for small-molecule compound identification},
  author={Li, Yuanyue and Kind, Tobias and Folz, Jacob and Vaniya, Arpana and Mehta, Sajjan Singh and Fiehn, Oliver},
  journal={Nature Methods},
  volume={18},
  number={12},
  pages={1524--1531},
  year={2021},
  publisher={Nature Publishing Group US New York}
}

@article{david2008structure,
  title={Structure determination from powder diffraction data},
  author={David, William IF and Shankland, Kenneth},
  journal={Foundations of Crystallography},
  volume={64},
  number={1},
  pages={52--64},
  year={2008},
  publisher={International Union of Crystallography}
}

@article{lai2025end,
  title={End-to-End Crystal Structure Prediction from Powder X-Ray Diffraction},
  author={Lai, Qingsi and Xu, Fanjie and Yao, Lin and Gao, Zhifeng and Liu, Siyuan and Wang, Hongshuai and Lu, Shuqi and He, Di and Wang, Liwei and Zhang, Linfeng and others},
  journal={Advanced Science},
  volume={12},
  number={8},
  pages={2410722},
  year={2025},
  publisher={Wiley Online Library}
}

@article{widdowson2022resolving,
  title={Resolving the data ambiguity for periodic crystals},
  author={Widdowson, Daniel and Kurlin, Vitaliy},
  journal={Advances in Neural Information Processing Systems},
  volume={35},
  pages={24625--24638},
  year={2022}
}

@article{dong2021deep,
  title={A deep convolutional neural network for real-time full profile analysis of big powder diffraction data},
  author={Dong, Hongyang and Butler, Keith T and Matras, Dorota and Price, Stephen WT and Odarchenko, Yaroslav and Khatry, Rahul and Thompson, Andrew and Middelkoop, Vesna and Jacques, Simon DM and Beale, Andrew M and others},
  journal={npj Computational Materials},
  volume={7},
  number={1},
  pages={74},
  year={2021},
  publisher={Nature Publishing Group UK London}
}

@article{chitturi2021automated,
  title={Automated prediction of lattice parameters from X-ray powder diffraction patterns},
  author={Chitturi, Sathya R and Ratner, Daniel and Walroth, Richard C and Thampy, Vivek and Reed, Evan J and Dunne, Mike and Tassone, Christopher J and Stone, Kevin H},
  journal={Applied Crystallography},
  volume={54},
  number={6},
  pages={1799--1810},
  year={2021},
  publisher={International Union of Crystallography}
}

@article{habershon2004powder,
  title={Powder diffraction indexing as a pattern recognition problem: a new approach for unit cell determination based on an artificial neural network},
  author={Habershon, Scott and Cheung, Eugene Y and Harris, Kenneth DM and Johnston, Roy L},
  journal={The Journal of Physical Chemistry A},
  volume={108},
  number={5},
  pages={711--716},
  year={2004},
  publisher={ACS Publications}
}

@article{zhang2024crystallographic,
  title={Crystallographic phase identifier of a convolutional self-attention neural network (CPICANN) on powder diffraction patterns},
  author={Zhang, Shouyang and Cao, Bin and Su, Tianhao and Wu, Yue and Feng, Zhenjie and Xiong, Jie and Zhang, T-Y},
  journal={IUCrJ},
  volume={11},
  number={4},
  pages={634--642},
  year={2024},
  publisher={International Union of Crystallography}
}

@article{schopmans2023neural,
  title={Neural networks trained on synthetically generated crystals can extract structural information from ICSD powder X-ray diffractograms},
  author={Schopmans, Henrik and Reiser, Patrick and Friederich, Pascal},
  journal={Digital Discovery},
  volume={2},
  number={5},
  pages={1414--1424},
  year={2023},
  publisher={Royal Society of Chemistry}
}

@inproceedings{bin2025simxrd,
  title={Simxrd-4m: Big simulated x-ray diffraction data and crystal symmetry classification benchmark},
  author={Bin, CAO and Liu, Yang and Zheng, Zinan and Tan, Ruifeng and Li, Jia and Zhang, Tong-yi},
  booktitle={The Thirteenth International Conference on Learning Representations},
  year={2025}
}

@article{oviedo2019fast,
  title={Fast and interpretable classification of small X-ray diffraction datasets using data augmentation and deep neural networks},
  author={Oviedo, Felipe and Ren, Zekun and Sun, Shijing and Settens, Charles and Liu, Zhe and Hartono, Noor Titan Putri and Ramasamy, Savitha and DeCost, Brian L and Tian, Siyu IP and Romano, Giuseppe and others},
  journal={npj Computational Materials},
  volume={5},
  number={1},
  pages={60},
  year={2019},
  publisher={Nature Publishing Group UK London}
}

%%%%%%%%%%%%%%%%%%%%%%%%%%%%%%%%%%%%%%%%%%%%%%%%%%
\clearpage
\section*{Supplementary Information}
\beginsupplement

\subsection{Selected Structures}

\clearpage
\subsection{XRD Representation}
\label{appndx:xrd_details}
We follow \citet{riesel2024crystal} and compute diffraction patterns from the structure factor contributions of each atomic site, in a differentiable way. Lattice parameters are converted to a real-space cell, the reciprocal lattice is derived, and all allowed Miller indices within the maximum scattering vector are generated. For each $(hkl)$, reciprocal distances and diffraction angles are calculated, elemental scattering factors are retrieved and weighted by site occupancies, and intensities are obtained by squaring the modulus of the summed structure factor with a Lorentz–polarization correction.

We calculate XRD peak profiles using the Pseudo-Voigt approximation, which models the peak shape as a linear combination of Gaussian and Lorentzian components:
\begin{equation*}
    pV(x) = \eta \, G(x) + (1 - \eta) \, L(x)
\end{equation*}
where $G(x)$ is the Gaussian function, $L(x)$ is the Lorentzian function, and $\eta \in [0,1]$ is the mixing parameter controlling the relative contributions.

\subsubsection{Gaussian and Lorentzian peak shapes}  
For a peak centered at $2\theta_0$, the Gaussian and Lorentzian components are given by:
\begin{align*}
    G(x) &= \exp\left[ -\frac{4 \ln 2 \, (x - 2\theta_0)^2}{H_G^2} \right], \\
    L(x) &= \frac{1}{1 + \frac{4 (x - 2\theta_0)^2}{H_L^2}},
\end{align*}
where $H_G$ and $H_L$ are the full widths at half maximum (FWHM) for the Gaussian and Lorentzian profiles, respectively.

\subsubsection{Caglioti parameters}  
In practice, peak broadening in XRD is described by the Caglioti relation:
\begin{equation*}
    H^2(2\theta) = U \, \tan^2\theta + V \, \tan\theta + W,
\end{equation*}
where $U$, $V$, and $W$ are the Caglioti parameters. This equation gives the squared FWHM as a function of diffraction angle, and is applied separately for the Gaussian and Lorentzian widths, i.e., $H_G(2\theta)$ and $H_L(2\theta)$. The parameters account for instrumental and sample-dependent broadening effects.

\subsubsection{Final pattern representation}  
We compute the total XRD pattern by summing $pV(x)$ contributions from all Bragg reflections over $2\theta \in [0^\circ, 90^\circ]$, and then discretize the intensity into bins of width $0.01^\circ$. We adopt Caglioti parameterss $U=0.1 , V=0.01, W=0.1$ and $\eta=0.1$. This produces a fixed-length $\text{xrd}$ vector, $\mathbf{x}$, of size $9000$ for each structure.

\subsection{Symmetry Projectors by Crystal Family}
\label{app:sym_op}

For each crystal family, the symmetry projector $\mathcal{P}$ maps the given lattice parameters 
$(a,b,c,\alpha,\beta,\gamma)$ to the symmetrized parameters consistent with the family:

\begin{align*}
    \mathcal{P}_{\mathrm{cubic}}(a,b,c,\alpha,\beta,\gamma)
    &= \big(\bar{a},\bar{a},\bar{a},\; 90^\circ,\; 90^\circ,\; 90^\circ\big), 
    & \bar{a} &= \frac{a+b+c}{3} \\
    \mathcal{P}_{\mathrm{hexagonal}}(a,b,c,\alpha,\beta,\gamma)
    &= \big(\bar{a},\bar{a},c,\; 90^\circ,\; 90^\circ,\; 120^\circ\big), 
    & \bar{a} &= \frac{a+b}{2} \\
    \mathcal{P}_{\mathrm{tetragonal}}(a,b,c,\alpha,\beta,\gamma)
    &= \big(\bar{a},\bar{a},c,\; 90^\circ,\; 90^\circ,\; 90^\circ\big), 
    & \bar{a} &= \frac{a+b}{2} \\
    \mathcal{P}_{\mathrm{orthorhombic}}(a,b,c,\alpha,\beta,\gamma)
    &= \big(a,b,c,\; 90^\circ,\; 90^\circ,\; 90^\circ\big) \\
    \mathcal{P}_{\mathrm{monoclinic}}(a,b,c,\alpha,\beta,\gamma)
    &= \big(a,b,c,\; 90^\circ,\; \beta,\; 90^\circ\big) \\
    \mathcal{P}_{\mathrm{triclinic}}(a,b,c,\alpha,\beta,\gamma)
    &= \big(a,b,c,\; \alpha,\; \beta,\; \gamma\big)
\end{align*}

\subsection{Implementation Details}
\label{appndx:imp_det}
Each optimization run was performed for a maximum of 30,000 iterations, with batch sizes of 16. For lattice distortions, the learning rates explored were between 0.0005 and 0.01. For coordinate distortions, the learning rates explored were between 1e-04 to 1e-06. The XRD parameters $U$, $V$, and $W$ (see Appendix Section~\ref{appndx:xrd_details}) were kept fixed at $[0.1, 0.01, 0.1]$, the same values used for generating the ground truth XRD patterns used for reference. Optimization was terminated early if the convergence tolerance of $10^{-4}$ in the loss function between 1000 steps was reached.

For lattice distorted structures, the lattice lengths and angles were optimized jointly while atomic coordinates were held fixed. For coordinate distortions, only coordinates were optimized. The loss functions considered included negative cosine similarity, negative entropy similarity, and mean squared error (MSE). In the MSE case, to ensure numerical stability, reflections were subject to physics-based clipping within Miller indices up to $[h,k,l]_\text{max}=6$.

For comparison with relaxation-based methods, structures were relaxed using CHGNet~\cite{deng2023chgnet}. For distorted coordinates, relaxation was done with the BFGS optimizer. For the distorted lattice case, relaxation was done with the FrechetCellFilter and BFGS optimizer. The relaxation process was terminated once the maximum atomic force ($f_\text{max}$) fell below 0.02~eV/Å.
\clearpage
\subsection{Correlations between AMD and XRD Similarity}
\label{appndx:corr}
\begin{table}[ht]
\centering
\caption{Pearson Correlation Between Average Minimum Distances (AMD) and XRD Similarity Metrics for the Different Noise Types and Levels.}
\label{tab:pearson_corr}
\begin{tabular}{c c c c c c}
\hline
\textbf{Noise Type} & \textbf{Noise Level} &  \makecell{\textbf{Cosine} \\ \textbf{Similarity}} & \textbf{MSE} & \makecell{\textbf{Entropy} \\ \textbf{Similarity}} & \makecell{\textbf{Cosine Similarity} \\ \textbf{+ Constraints}} \\
\hline
\multirow{5}{*}{Lattice} 
 & 0.01 & 0.76 & 0.59 & 0.31 & \textbf{0.79}\\
 & 0.05 & 0.43 & 0.4 & 0.53 & \textbf{0.83} \\
 & 0.1  & 0.36 & 0.27 & 0.39 & \textbf{0.76} \\
 & 0.15 & 0.16 & 0.1 & 0.23 & \textbf{0.72}  \\
 & 0.2  & 0.05 & 0.1 & 0.13 & \textbf{0.65} \\
\hline
\multirow{5}{*}{Coordinates} 
 & 0.001 & -0.64 & \textbf{0.23} & -0.6 & -\\
 & 0.005 & -0.72 & \textbf{0.27} & -0.55 & -\\
 & 0.01  &  -0.64 & \textbf{0.46} & -0.64 & -\\
 & 0.05  & -0.02 & \textbf{0.51} & -0.27 & -\\
 & 0.1   & \textbf{0.52} & 0.48 & 0.43 & -\\
\hline
\end{tabular}
\end{table}

\begin{table}[ht]
\centering
\caption{Spearman Correlation Between Average Minimum Distances (AMD) and XRD Similarity Metrics for the Different Noise Types and Levels.}
\label{tab:spear_corr}
\begin{tabular}{c c c c c c}
\hline
\textbf{Noise Type} & \textbf{Noise Level} &  \makecell{\textbf{Cosine} \\ \textbf{Similarity}} & \textbf{MSE} & \makecell{\textbf{Entropy} \\ \textbf{Similarity}} & \makecell{\textbf{Cosine Similarity} \\ \textbf{+ Constraints}} \\
\hline
\multirow{5}{*}{Lattice} 
 & 0.01 & \textbf{0.93} & 0.61 & 0.65 & 0.65 \\
 & 0.05 & 0.47 & 0.43 & 0.58 & \textbf{0.89} \\
 & 0.1  & 0.39 & 0.29 & 0.39 & \textbf{0.77} \\
 & 0.15 & 0.18 & 0.1 & 0.25 & \textbf{0.67}  \\
 & 0.2  & 0.3 & 0.13 & 0.12  & \textbf{0.55} \\
\hline
\multirow{5}{*}{Coordinates} 
 & 0.001 & -0.67 & \textbf{0.74} & -0.64 & -\\
 & 0.005 & -0.64 & \textbf{0.78} & -0.51 & -\\
 & 0.01  & -0.55 & \textbf{0.84} & -0.62 & -\\
 & 0.05  & -0.06 & \textbf{0.63} & -0.46 & -\\
 & 0.1   & 0.5 & \textbf{0.52} & 0.28 & -\\
\hline
\end{tabular}
\end{table}

\subsection{2D Landscape of XRD-Based Optimization}
\vspace*{-0.5em}

\clearpage
\vspace*{-2em}

\subsection{XRD Similarity vs. Structural Agreement}

\subsection{2D Landscape of Energy-Based Optimization}
\vspace*{-0.5em}

%%%%%%%%%%%%%%%%%%%%%%%%%%%%%%%%%%%%%%%%%%%%%%%%%%

\end{document}